\documentclass[sigconf, nonacm]{acmart}

\newcommand\vldbdoi{XX.XX/XXX.XX}

\newcommand\vldbavailabilityurl{https://github.com/Wangyibo321/WAter}
\usepackage[linesnumbered,ruled,vlined]{algorithm2e} 
\usepackage{epstopdf}
\usepackage{amsmath}
\DeclareMathOperator*{\argmax}{arg\,max}
\usepackage{multicol}
\usepackage{url}
\usepackage{enumerate}
\usepackage[labelfont=bf]{caption}
\usepackage{xcolor}
\usepackage{color, colortbl}
\usepackage{hyperref}
\usepackage{algpseudocode}
\usepackage{tabularx}

\usepackage{subfig}
\usepackage{graphicx}
\usepackage{pifont}
\usepackage{enumitem} 
\usepackage{balance}

\usepackage{multirow}
\usepackage{diagbox}
\usepackage{CJKutf8} 

\usepackage{boldline}

\usepackage[table]{xcolor}
\definecolor{headergray}{HTML}{B0B3B2}
\definecolor{rowgray}{HTML}{D4D4D4}

\usepackage{soul} % underline

\usepackage{bibunits}

\newcommand{\system}{{\textsc{WAter}}}

\begin{document}
\title{\system: A Workload-Adaptive Knob Tuning System based on Workload Compression}

%%
%% The "author" command and its associated commands are used to define the authors and their affiliations.

\author{Yibo Wang}
\orcid{0009-0005-1971-3398}
\affiliation{%
  \institution{Purdue University}
  \city{West Lafayette, Indiana}
  \country{USA}
}
\email{wang7342@purdue.edu}

\author{Jiale Lao}
\orcid{0009-0003-1144-5152}
\affiliation{%
  \institution{Cornell University}
  \city{Ithaca, New York}
  \country{USA}
}
\email{jiale@cs.cornell.edu}

\author{Chen Zhang}
\affiliation{%
  \institution{Sichuan University}
  \city{Chengdu}
  \country{China}
}

\email{zhangchen2@stu.scu.edu.cn}
\author{Cehua Yang}
% \orcid{0009-0004-4285-5696}
\affiliation{%
  \institution{Sichuan University}
  \city{Chengdu}
  \country{China}
}
\email{yangcehua@stu.scu.edu.cn}

% \author{Yuanchun Zhou}
% \orcid{0000-0003-2144-1131}
% \affiliation{%
%   \institution{Computer Network Information Center, Chinese Academy of Sciences}
%   \city{Beijing}
%   \country{China}
% }
% \email{zyc@cnic.cn}

\author{Jianguo Wang}
\orcid{0000-0002-3039-1175}
\affiliation{%
  \institution{Purdue University}
  \city{West Lafayette, Indiana}
  \country{USA}
}
\email{csjgwang@purdue.edu}

\author{Mingjie Tang}
\orcid{0000-0002-8893-4574}
% \authornote{The corresponding author.}
\affiliation{%
  \institution{Sichuan University}
  \city{Chengdu}
  \country{China}
}
\email{tangrock@gmail.com}

\begin{abstract}
Selecting appropriate values for the configurable parameters of Database Management Systems (DBMS) to improve performance is a significant challenge. \textcolor{black}{Recent machine learning (ML)-based tuning systems have shown strong potential, but their practical adoption is often limited by the high tuning cost. This cost arises from two main factors: (1) the system needs to evaluate a large number of configurations to identify a satisfactory one, and (2) for each configuration, the system must execute the entire target workload on the DBMS, which is both time-consuming and resource-intensive. Existing studies have primarily addressed the first factor by improving sample efficiency, that is, by reducing the number of configurations that must be evaluated. However, the second factor, improving \textit{runtime efficiency} by reducing the time required for each evaluation, has received limited attention and remains an underexplored direction.} 

We develop \system, a runtime-efficient and workload-adaptive tuning system that finds near-optimal configurations at \textit{a fraction of the tuning cost} compared with state-of-the-art methods. Instead of repeatedly replaying the \textcolor{black}{entire target} workload, we divide the tuning process into multiple time slices and evaluate only a small subset of representative queries from the workload in each slice. Different subsets are evaluated across slices, and a runtime profile is used to dynamically identify more representative subsets for evaluation in subsequent slices. At the end of each time slice, the most promising configurations are selected and evaluated on the original workload to measure their actual performance. Technically, we design a query-level metric and use a greedy algorithm that continually refines the query subset (e.g., removing uninformative queries and adding promising ones) as the tuning progresses. \textcolor{black}{We then develop a hybrid scoring mechanism, built upon a global surrogate model, to balance exploitation and exploration and to recommend promising configurations for evaluation on the entire workload. Finally, we evaluate \system ~across different workloads and compare it with state-of-the-art approaches. \system ~identifies the best-performing configurations with up to $\mathbf{73.5\%}$ less tuning time and achieves up to $\mathbf{16.2\%}$ higher performance than the best-performing alternative. We also demonstrate \system’s robustness across different hardware platforms and optimizers, as well as its scalability across database sizes.} 
\end{abstract}

\maketitle

% %%% do not modify the following VLDB block %%
% %%% VLDB block start %%%
% \pagestyle{\vldbpagestyle}
% \begingroup\small\noindent\raggedright\textbf{PVLDB Reference Format:}\\
% \vldbauthors. \vldbtitle. PVLDB, \vldbvolume(\vldbissue): \vldbpages, \vldbyear.\\
% \href{https://doi.org/\vldbdoi}{doi:\vldbdoi}
% \endgroup
% \begingroup
% \renewcommand\thefootnote{}\footnote{\noindent
% This work is licensed under the Creative Commons BY-NC-ND 4.0 International License. Visit \url{https://creativecommons.org/licenses/by-nc-nd/4.0/} to view a copy of this license. For any use beyond those covered by this license, obtain permission by emailing \href{mailto:info@vldb.org}{info@vldb.org}. Copyright is held by the owner/author(s). Publication rights licensed to the VLDB Endowment. \\
% \raggedright Proceedings of the VLDB Endowment, Vol. \vldbvolume, No. \vldbissue\ %
% ISSN 2150-8097. \\
% \href{https://doi.org/\vldbdoi}{doi:\vldbdoi} \\
% }\addtocounter{footnote}{-1}\endgroup

\ifdefempty{\vldbavailabilityurl}{}{
\vspace{.3cm}
\begingroup\small\noindent\raggedright\textbf{PVLDB Artifact Availability:}\\
The source code, data, and/or other artifacts have been made available at \url{\vldbavailabilityurl}.
\endgroup
}

\section{Introduction}

\label{sec: intro}

{Database management systems (DBMSs) rely on many configuration parameters (i.e., knobs) to control their behavior \cite{pavlo2017self}. Tuning these knobs is crucial for achieving high performance~\cite{inquiry}.} Conventionally, these knobs are adjusted manually by database administrators (DBAs), involving extensive workload, system, and hardware analysis. However, DBAs encounter substantial difficulties identifying promising configurations for a specific workload due to the high dimensionality of the configuration space, where each knob can have continuous or discrete values (heterogeneity). This challenge becomes even more pronounced in the cloud, where the underlying hardware resources can vary significantly across DBMS instances.

Recent works focus on using Machine Learning (ML) techniques to automate knob tuning to reduce the manual tuning efforts, and have shown promising results \cite{ituned,black_or_white,cgptuner,gptuner,restune,llamatune,dbbert,onlinetune,ddpg1,ddpg2,udo,opadviser}. These ML-based tuning
systems iteratively select a configuration using a tuner, balancing between the exploration of unseen regions and the exploitation of known space. {{The selected configurations are then evaluated by executing the target workload on the DBMS.}} Since it is challenging to explore the high-dimensional and heterogeneous search space, many techniques are proposed to explore the space efficiently, such as search space pruning \cite{gptuner,opadviser} and transfer learning \cite{ottertune,qtune,restune,cgptuner}.

% \vspace{-1em}

\begin{figure}[h]
    \centering
    \includegraphics[width=0.9\linewidth]{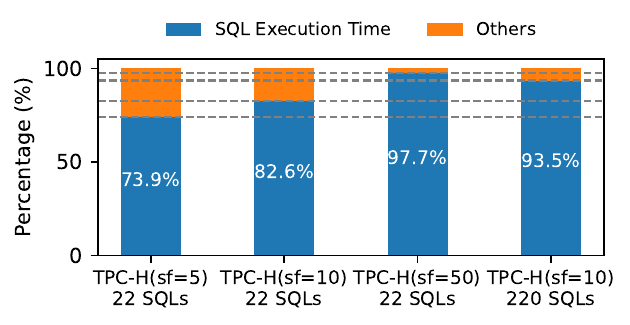}
    \vspace{-1em}
    \caption{Tuning Time Breakdown (Percentage)}
    \label{fig:tuning_time_breakdown}
    % \vspace{-1em}
\end{figure}

Although state-of-the-art systems reduce the required iterations to only hundreds to identify ideal configurations, the tuning cost is \textit{still high} because it takes a long time to execute the workload in \textit{each iteration}. For example, {in our experiment in Figure \ref{fig:different_scale_factor}}, it takes 10 minutes to execute the 22 queries in the TPC-H benchmark with a scale factor of 50, \textcolor{black}{leading to about 17 hours of optimization for 100 valid iterations}.  Figure~\ref{fig:tuning_time_breakdown} shows the breakdown of the tuning time of a state-of-the-art method \cite{gptuner} for TPC-H benchmark under different scale factors. Notably, more than 70\% of tuning time is spent on executing the target workload on DBMS, and this becomes more pronounced (e.g., more than 97\%) as the data size increases or the workload becomes more complex, an observation similar to previous work \cite{inquiry}.

Therefore, we argue that \textit{it is important to reduce the workload execution time while keeping the tuning effective}, given that the major tuning costs come from substantial workload execution time, a factor overlooked by prior research. 
In this paper, we propose a new concept of \textit{runtime efficiency}, which refers to minimizing the workload execution time in each tuning iteration and thus achieving the overall minimum tuning time. This approach is compatible with previous works focusing on decreasing the number of tuning iterations, but goes one step further by trying to reduce the running time of each iteration.

However, achieving runtime efficiency of knob tuning presents several challenges. \textit{\textbf{C1. It is non-trivial to reduce the workload execution time.}} There are two possible ways to cut down the workload execution time: one is to decrease the volume of the target database and the other is to reduce the number of queries in the workload (workload compression). Decreasing the data volume by sampling a subset of data can severely affect the performance. Because knob tuning is highly sensitive to the underlying data, reducing the data size is likely to change the performance bottleneck and thus mislead the tuning process. In contrast, workload compression \cite{2002,2003,gsum,isum,wred,query2vec} presents a promising approach. It aims to identify a substitute query subset that approximates the runtime behavior of the DBMS under the full workload, without significantly degrading the performance of workload-driven tasks like index tuning \cite{2002}. 

Unfortunately, \textit{\textbf{C2. It is challenging to obtain a compressed workload that is truly representative for the specific task of knob tuning.}} 
While there has been substantial work on workload compression~\cite{gsum, query2vec, isum, 2002, wred, refactoring}, these techniques are not effective in our context because they were originally designed for index tuning~\cite{2002, isum, wred, refactoring, index_survey}, which is a fundamentally different problem that focuses on selecting table columns for index construction. These methods often rely on query-level features such as shared ``indexable columns'', which do not translate to the knob tuning task. Index-agnostic approaches, including random sampling and GSUM~\cite{gsum}, only use generic workload information and therefore perform sub-optimally for specialized tasks (e.g., index tuning or knob tuning), as confirmed by our experimental results in Section~\ref{subsec:perf-comparison}. Selecting a representative subset of queries as the compressed workload for knob tuning remains an open challenge.

\textit{\textbf{C3: Knob tuning introduces new challenges when applying workload compression, as good performance on a subset of the workload does not necessarily imply good performance on the full workload.}} When tuning one subset, we only evaluate the configurations on this subset. However, a good configuration for this subset does not necessarily perform well on the original workload, and may even lead to performance degradation. Moreover, even if the tuning is guided by advanced optimization algorithms ~\cite{bo, rl}, there is no guarantee that the configurations recommended later are better than the previous ones (not monotonic). Therefore it is infeasible to simply evaluate configurations from later iterations on the original workload. We need a sophisticated mechanism to identify well-performing configurations for the entire workload without verifying every proposed option, as doing so is exhausting and would negate the benefits of workload compression. Challenges remain regarding how to determine whether a configuration is worth evaluating, how to trade off between subset tuning and configuration verification, whether the subset should be dynamically updated and if so, how to achieve that.

\noindent \textbf{Our Approach.} To address these challenges, we develop \system, a runtime-efficient and workload-adaptive tuning system, and it identifies near-optimal configurations at a fraction of the tuning time compared to state-of-the-art methods. The key observation of \system~ is that aforementioned limitations of existing approaches (\textit{\textbf{C1}} and \textit{\textbf{C2}}) are rooted in the intractable difficulty to find a perfect subset in one try. Differently, \system~ starts with an imperfect subset and continually refine it based on runtime profile on the fly (Section \ref{sec: runtime-adaptive}). Instead of replaying the whole workload or a fixed subset repeatedly, we divide the tuning process into many time slices and evaluate different subsets at different time slices (Section \ref{sec: formulation}). To continually refine the subset as the tuning proceeds, we carefully design a runtime metric to measure the representativity of a subset to its original workload (Section \ref{sec: representative}), and propose to use a greedy algorithm to optimize this metric (Section \ref{sec: runtime-adaptive}).
Moreover, to mitigate the overhead of switching between tuning different subsets, we develop a history reuse mechanism for efficient subset tuning (Section \ref{sec: history-reuse}). Regarding \textit{\textbf{C3}}, we design heuristic-based rules to prune unpromising configurations (e.g., configurations perform significantly worse than the default configuration are discarded). After pruning, we propose a hybrid scoring mechanism to score and rank configurations, only verifying the most promising configurations on the original workload. The scoring mechanism is based on a global surrogate model, predicting the performance as well as the uncertainty of the prediction for configurations to balance between exploration and exploitation (Section \ref{sec: configuration}). Finally, we conduct extensive experiments to evaluate \system's effectiveness, robustness and scalability.

\noindent \textbf{Experimental Overview.}  
Our extensive experiments demonstrate \system's decisive advantages over state-of-the-art tuners across multiple OLAP benchmarks. On average, \system\ finds optimal configurations \textbf{$4.2\times$} faster while discovering superior solutions that yield up to \textbf{$16.2\%$} better final performance, with time-to-optimal speedups reaching a remarkable \textbf{$12.9\times$} on complex workloads. This state-of-the-art performance is proven to be robust across different hardware, optimizers, and larger database scales where \system's runtime efficiency provides the greatest benefit. A detailed ablation study confirms the criticality of each of our core components, while a cost analysis reveals that \system's efficiency stems from drastically reducing the dominant cost of workload evaluation time.

\noindent \textbf{Contributions.} Our contributions are as follows. (1) We develop \system, a \textit{runtime-efficient} knob tuning system that identifies near-optimal configurations at a fraction of the tuning time compared to state-of-the-art methods. (2) We introduce a new paradigm that applies workload compression to enhance the knob tuning process and identify the associated technical challenges. (3) We develop a set of techniques to address these challenges, including: a time-slicing design that partitions the tuning process into multiple time intervals and evaluates different query subsets across them (Section \ref{sec: overview}); an adaptive mechanism that incrementally refines the query subset using a greedy algorithm to make it increasingly representative of the complete workload based on runtime feedback (Section \ref{sec: workload compression}); a history reuse mechanism that minimizes the overhead of switching between query subsets (Section \ref{sec: history-reuse}); and a hybrid scoring algorithm that selects only the most promising configurations for validation (Section \ref{sec: configuration}).

\section{Background and Related Work}
\label{sec: background}

\subsection{Database Knob Tuning}
\label{sec: knobtuning}

\textbf{Database Tuning Problem.} We formulate database knob tuning as an optimization problem. Given a \textit{target workload} $\mathbf{W}$ and the \textit{configuration space} $\boldsymbol{\Theta}$, the performance metric is given by an \textit{objective function} $f_\mathbf{W}:\boldsymbol{\Theta}\rightarrow 
 \mathbb{R}$, that projects each configuration to a value of the performance metric (e.g., latency or throughput). Database knob tuning aims to find a configuration $\theta^{*}\in\boldsymbol{\Theta}$, where
\begin{equation}
    \boldsymbol{\theta}^{*}= \argmax_{\theta \in \boldsymbol{\Theta}} f_\mathbf{W}(\theta)
    \label{eq: 1}
\end{equation}

{Finding an optimal database configuration is challenging due to the vast configuration space}. Such difficulty goes beyond the capability of even the best human experts, so database community turns to ML-based automatic tuning methods.

\begin{figure}[h]
    \centering

    \includegraphics[scale=0.35]{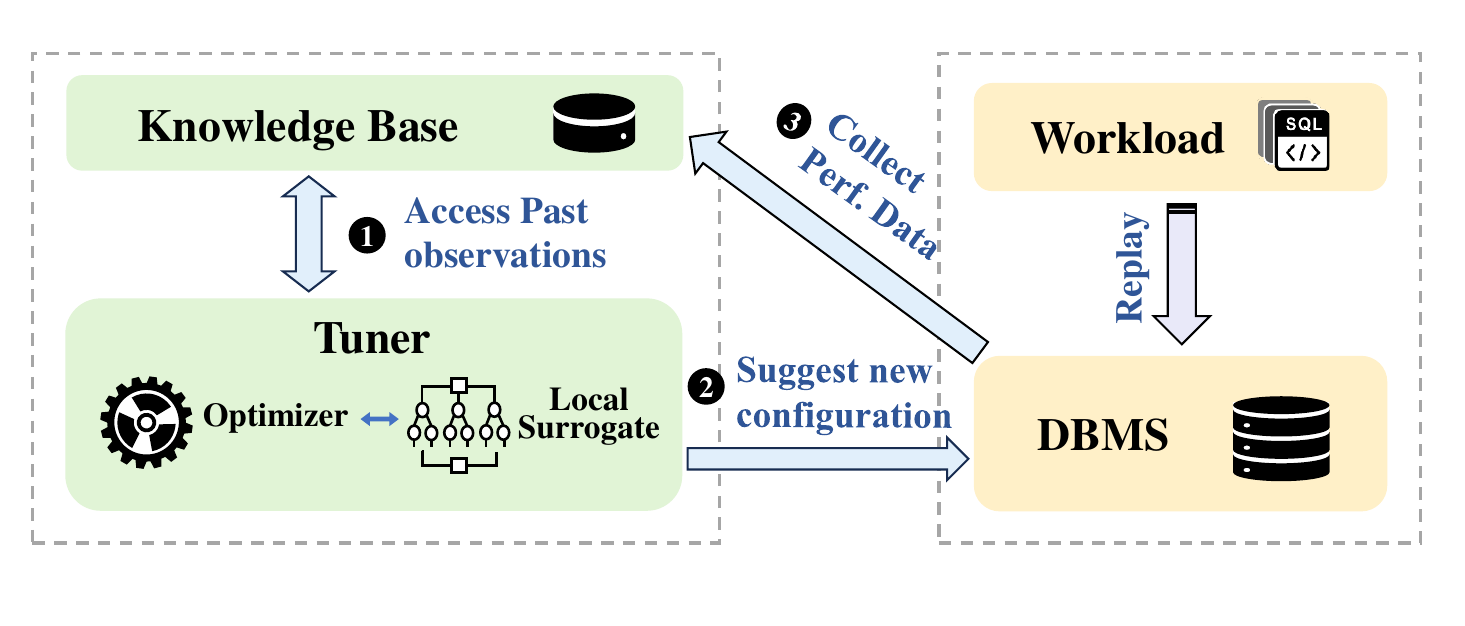}
    \vspace{-3em}
    \caption{Overview of Knob Tuning Paradigm}
    
    \label{fig: knob_tuning}
     % \vspace{-1em}
\end{figure}

\noindent \textbf{ML-based Knob Tuning.} Recently, ML-based approaches have demonstrated promising results, achieving better performance than human DBAs as well as static rule-based tuning tools \cite{pgtune,probiblistic}. Moreover, ML-based approaches are automatic and can adapt well to a variety of workloads and hardware configurations. Figure \ref{fig: knob_tuning} presents the paradigm of the ML-based knob tuning framework which mainly contains \textit{(i)} a \textit{tuner} that suggests a configuration over a given search space to improve the pre-defined performance metrics, and \textit{(ii)} a \textit{DBMS instance} that runs the workload under the proposed configuration to obtain the performance metric. The knowledge base $\mathcal{D}=\{\theta_i, f_\mathbf{W}(\theta_i)\}$ is an optional component which records all previously evaluated configurations, and updates every time a new evaluation is conducted.
These systems can be broadly classified into two main categories based on the techniques used in the \textit{tuner}: Bayesian Optimization (BO)-based \cite{bo} and Reinforcement Learning (RL)-based \cite{rl}.

% \vspace{-0.5em}

% \vspace{-0.5em}

\noindent \labelitemi~ \textbf{RL-based.} RL-based methods explore the configuration space in a trial-and-error manner. The agent (e.g., a neural network) iteratively tries new configurations and learns from the rewards (e.g., performance improvement or degradation) obtained from the environment (e.g., DBMS). Deep Deterministic Policy Gradient (DDPG)~\cite{ddpg} is the most popular RL algorithm adopted in knob tuning \cite{ddpg1, qtune, ddpg2}, as DDPG can work over a continuous space.
% instead of setting a knob from a finite set of predefined values.

\noindent \labelitemi~ \textbf{BO-based.} BO-based methods \cite{cgptuner,gptuner,ituned,restune,ottertune,onlinetune,gptuner-demo} model the tuning as a black-box optimization problem.
% Bayesian optimization is a model-based approach that aims to identify the optimum of an unknown function $f$ that is expensive to evaluate, while minimizing the number of samples taken.
BO consists of two main components: (1) \textit{surrogate model} is an ML model to approximate the objective function $f_\mathbf{W}$, given a set of observations $\{\theta_i, f(\theta_i)\}$. 
It provides both a prediction and the confidence of the prediction for an unseen configuration. 
% \textit{Surrogate model} is updated each time a new configuration is evaluated.
(2) \textit{acquisition function} uses the surrogate model's outputs to choose which candidate point to evaluate next, balancing between exploitation and exploration.

\noindent\textcolor{black}{\noindent \labelitemi~ \textbf{Tuning Frameworks.} Frameworks such as MLOS \cite{mlos} do not directly improve tuning efficiency, but serve to bridge the gap among benchmarking, experimentation, and optimization. In contrast, other frameworks like OtterTune \cite{ottertune} and LlamaTune \cite{llamatune} are designed to enhance tuning efficiency. These frameworks focus on sample efficiency and are orthogonal to our work, which focuses on runtime efficiency. They can co-exist with \system\ to further enhance the performance of existing optimizers.
}

\noindent \textbf{Performance Comparison.} According to \cite{ea}, RL-based methods require more iterations to work well due to the complexity of the neural networks used. The majority of previous works use BO-based methods, and \cite{ea} concluded that the best performing optimizer was Sequential Model-based Algorithm Configuration (SMAC \cite{smac3}), since it is efficient in modeling the heterogeneous search space. With the recent advent of Large Language Model (LLM),  \textsc{GPTuner}~\cite{gptuner} uses LLM to read manuals and constructs structured knowledge to guide the BO-based tuning process. \textit{We regard \textsc{GPTuner} and SMAC as the current state-of-the-art methods with and without text as inputs. We integrate \system \ with these methods.}

% \vspace{-1em}
\subsection{Workload Compression}
\label{sec: background workload}

Workload compression is first studied in \cite{2002}. Given a workload $\textbf{W}$, it aims to find a SQL subset $\textbf{W'}$ ($\textbf{W'}$ has fewer queries and each query comes from $\textbf{W}$), such that the workload execution cost is reduced (fewer queries to execute for knob tuning, or fewer columns to consider for index tuning), and the tuning performance does not degrade too much at the same time. However, the performance degradation is inevitable in practice. Existing works essentially trade performance for runtime efficiency \cite{wred,isum,gsum,query2vec}. The primary aim of our work differs significantly from previous works. Instead of trading performance for runtime efficiency, our approach can achieve superior performance compared to tuning the original workload within the same time budget, as reduced iteration costs allow for more thorough exploration of the configuration space. A detailed formulation of the problem and its underlying intuition can be found in Section \ref{sec: formulation}.

There are both generic and indexing-aware workload compression techniques in the literature. GSUM \cite{gsum} is a recent generic workload compression system that maximizes the coverage of features (e.g., columns contained) of the workload while ensuring that the compressed workload remains representative (i.e., having similar distribution to that of the entire workload). For indexing-aware compression, ISUM \cite{isum} selects queries greedily based on their potential to reduce the costs and the similarity between queries, and the two metrics are computed using indexing-specific featurization.
The most recent work, \textsc{Wred} \cite{wred}, rewrites each query in the original workload to eliminate columns and table expressions that are unlikely to benefit from indexes. These methods compress the workload in a single step, lacking further refinement. More importantly, they require manual feature engineering of queries, and some even require indexing-specific features, making these methods not applicable to knob tuning.
In contrast, \system \ focuses on knob tuning that seamlessly integrates workload compression through the entire tuning process, continuously refining the subset. Additionally, \system \ does not rely on any form of featurization; instead, it selects queries based on runtime statistics, allowing it to handle any executable query. In comparison, methods like \textsc{Wred} are unable to handle 19 out of 99 queries from TPC-DS that its parser cannot process.

\textcolor{black}{A recent concurrent work, SCompression \cite{scompression}, also addresses the high cost of workload execution and 
% Recognizing that prior efforts have focused on compressing Online Analytical Processing (OLAP) workloads, 
targets Online Transaction Processing (OLTP) workloads. It uses time-slices as the compression unit to preserve the inherent concurrency and temporal relationships of the original workload, and performs a one-time, static compression to generate a fixed workload that is used for the entire tuning process. 
% However, our experiments in Section \ref{subsec:perf-comparison} reveal that workload compression for knob tuning is non-trivial, and existing methods like GSUM \cite{gsum} exhibit suboptimal performance in this context.
In contrast, our approach in \system\ is fundamentally different in several key aspects. First, \system\ is designed specifically for OLAP workloads, where it operates by selecting a representative subset of individual queries from the entire workload. Our experiments in Section~\ref{subsec:perf-comparison} show that compressing OLAP workloads for knob tuning is a non-trivial task. Existing OLAP compression techniques, such as GSUM~\cite{gsum} and random sampling, yield poor performance in this setting. More importantly, unlike SCompression's static approach, \system\ introduces a dynamic and adaptive compression strategy that continually refines its selected query subset throughout the tuning process based on an evolving runtime profile. This adaptability prevents the tuning process from being misled by a fixed suboptimal subset, ensuring more robust optimization results. We summarize the main differences between them in Table \ref{tab: diff}.}

\begin{table}[h]
\caption{\textcolor{black}{Main Difference between WAter and SCompression}}
\centering % Centers the table on the page
\begin{tabular}{|c|c|c|c|}
\hline
\rowcolor{headergray} % Colors the entire header row
 & {Target Workload} & {Comp. Unit} & {Strategy} \\
\hline
\cellcolor{rowgray} WAter & OLAP & Query & Dynamic \\
\hline
\cellcolor{rowgray} SCompression & OLTP & Time slice & Static \\
\hline
\end{tabular}
\label{tab: diff}%
\end{table}

Some works on training data collection also involve sampling a SQL subset from the original workload. However, they focus on different application scenarios. From the perspective of model training, these works either aim to minimize the cost to obtain a labeled training dataset \cite{activelin,hitthegym} or select the most valuable training data (queries) \cite{shifting} for a learned database component (e.g., learned cost estimators) effectively. Moreover, in contrast to knob tuning, their target workload is typically a streaming query workload produced in the online scenario, rather than a fixed set of queries.

\section{Motivation}

In this section, we discuss the motivations behind the design and implementation of \system~ as well as how this paper is structured.

\noindent \textbf{\color{black}M1: \color{black} The search space for knob tuning is extremely large yet underexplored.}
The search space of knob tuning is extremely large due to: (1) \textit{the large number of knobs that require tuning}, and (2) \textit{the wide value range for each knob}. For example, PostgreSQL v14.9 has 346 knobs, and some most frequently tuned knobs like \texttt{shared\_buffers} range from 0.125 MB to 8192 GB, and \texttt{random\_page\_cost} can be set to any real value between $0$ and $1.79769\times10^{308}$. Moreover, some methods \cite{cgptuner,onlinetune} even add contextual information (e.g., workload feature) into the space which could further expand it. In the literature, it is commonly assumed that the number of evaluations required to find an optimum is proportional to the size of the search space \cite{nfl}. However, existing ML-based tuning methods only conduct hundreds to at most thousands of samplings and evaluations \cite{gptuner,cgptuner,restune,opadviser,llamatune,rover}, which is very sparse in such a colossal search space. The exploration of the search space is insufficient, and we need to explore it more thoroughly to identify better configurations.

\begin{figure}[tbp]
    \centering

    \includegraphics[scale=0.5]{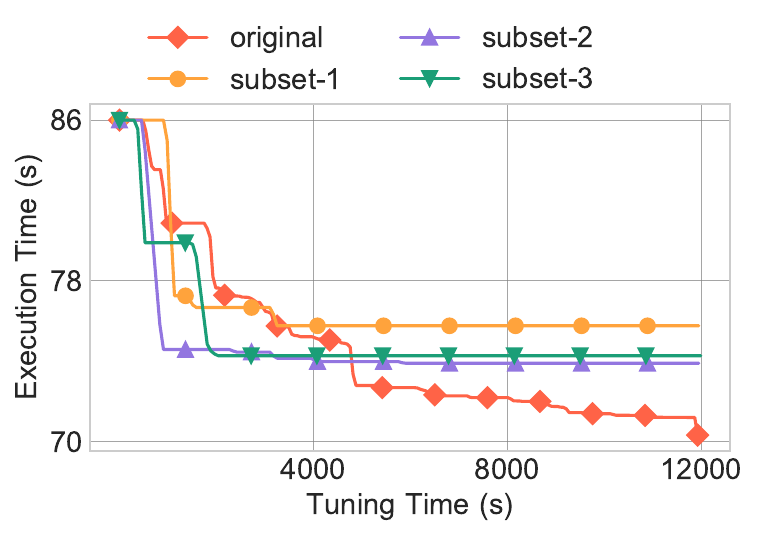}
    \vspace{-1em}
    \caption{Tuning Subsets VS Tuning the Original Workload}
    \label{fig: show_random}
% \vspace{-2em}
\end{figure}

\noindent \textbf{\color{black}M2: Under-exploration stems from high workload execution time, workload compression presents a promising method to reduce the costs.}
As discussed in Section \ref{sec: knobtuning}, evaluating a configuration requires executing the target workload, with each workload execution taking minutes or more. Such high costs greatly limit the number of configurations to try. A naive approach to mitigate \textbf{M1} involves sampling a small subset of queries from the original workload. By executing fewer queries, we decrease the workload execution time, allowing for exploration of a larger portion of the search space within a given time budget. We conduct an experimental study for this idea by randomly sampling 3 subsets of 26 queries from TPC-DS's 88 queries, using \textsc{GPTuner} \cite{gptuner} as the optimizer for its efficiency. Whenever a proposed configuration outperforms the default configuration on the subset, this configuration is immediately evaluated on the original workload to obtain real performance. We also use \textsc{GPTuner} to tune the original workload directly as a comparison. Figure \ref{fig: show_random} shows the latency of the best configuration found (y axis) as a function of optimization time (x axis).  It is worth noting that \textit{tuning a subset can make the tuner produce well-performing configurations with much less time.} 
The reason is that reduced execution time enables more configuration evaluations, improving exploration and increasing the likelihood of finding optimal solutions.

\noindent \textbf{\color{black}M3: \color{black} Identifying a representative subset is important but very challenging.}
From Figure \ref{fig: show_random}, we find that different subsets can lead to different optimization results, and a bad subset can make the optimization stuck in local optima and fail to find better configurations even after a long tuning time. An interpretation could be that, tuning a subset essentially involves optimizing an alternative objective function that approximates the real objective function of the entire workload, and the similarity between the objective functions of query subsets and the objective function of the original workload differs greatly for different subsets. A more representative subset can result in faster and more thorough optimizations, while a bad subset could even mislead the process. Selecting a good subset is critical for the end-to-end tuning performance, but unfortunately, we do not even have a method to quantify the representativity of a subset to its original workload in the context of knob tuning. 

\noindent \textbf{M4: It is nearly impossible to find a perfect subset in a single attempt, but we can continually refine the subset based on the evolving runtime profile.} Knob tuning is such a complex problem which involves almost all aspects of a DBMS, including resource management, background process management, query optimization and execution, and so on \cite{survey}. Therefore, it is almost impossible to identify a perfect subset at the beginning in a single attempt, just based on the workload information. Given the iterative nature of knob tuning, runtime statistics are accumulated incrementally throughout the tuning process. So it is reasonable to select a good but not perfect subset as the starting point, and then we continually refine this subset based on the evolving runtime profile.

\noindent\textbf{Outline.} To alleviate the under-exploration issue caused by costly workload execution (\textbf{M1}), we propose to just tune a subset and find this approach promising (\textbf{M2}). Although we find that identifying a representative subset is crucial for effective tuning, this process is challenging due to the absence of a metric to quantify the representativity of a subset to its original workload (\textbf{M3}). Moreover, given the complexity of knob tuning, it is too difficult to identify a representative subset in a single attempt (\textbf{M4}). Therefore, we make the following technical contributions. To handle \textbf{M3}, we propose (1) a \textit{representativity} metric based on runtime profile in Section \ref{sec: representative}, and (2) use a greedy algorithm to dynamically compress the workload in Section \ref{sec: runtime-adaptive}.
% , and (2) an LLM-based algorithm for cold-start scenario in Section \ref{sec: cold-start}
Based on \textbf{M4}, we develop (3) a workload-adaptive knob tuning framework that periodically updates the subset in Section \ref{sec: overview}. Moreover, it (4) reuses runtime statistics for efficient subset tuning in Section \ref{sec: history-reuse}, and (5) prunes, scores and ranks the proposed configurations for verification in Section \ref{sec: configuration}.

% \vspace{-1em}
\section{System Overview}
\label{sec: overview}
\system \ is a workload-adaptive knob tuning system that speeds up the tuning process by reducing workload execution time using workload runtime profile. The high-level idea is that instead of repeatedly executing the entire complex workload, we split the tuning process into a series of short time slices and evaluate only a small subset of SQL queries in each. \textcolor{black}{A time slice is a tuning cycle, where \system \ selects a representative SQL subset (Section \ref{sec: workload compression}), tunes the subset to obtain configurations (Section \ref{sec: history-reuse}) and finally evaluates promising configurations over the original workload (Section \ref{sec: configuration}).} Different subsets are selected in different time slices, and we continuously refine the subset based on evolving runtime profile. 

\begin{figure}[tb]
    \centering
    % \hspace*{-3cm}
    \includegraphics[scale=0.45]{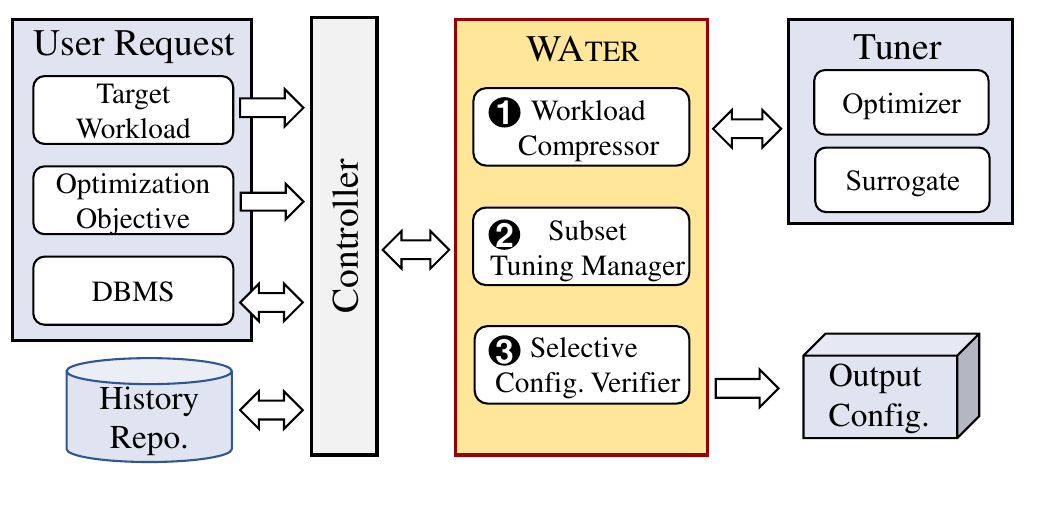}
    \vspace{-1em}
    \caption{Overview of the Components in the \system \ System}
    \label{fig: architecture}
    \vspace{-1em}
\end{figure}

\noindent \textit{\underline{Architecture.}} Figure \ref{fig: architecture} presents an overview of the architecture of \system. On the client side, the user provides the target workload, optimization objective (e.g., throughput or latency) and the DBMS to tune. The \textit{controller} deploys new configurations on DBMS, executes a set of queries, and collects performance metrics. \system \ interacts with the \textit{controller} to request query execution under specified configurations, gather the resulting execution data, and store it in the \textit{history repository}. \system \ contains three modules corresponding to the three steps in a time slice. 
First, the \textit{workload compressor} uses the runtime profile to select a representative subset of queries from the target workload. Second, the \textit{subset tuning manager} designates this SQL subset as the target workload for the current time slice and reuses existing tuning history to bootstrap the \textit{tuner}'s \textit{surrogate}, thereby enabling efficient subset tuning. Third, the \textit{selective configuration verifier} prunes, ranks, and selects configurations proposed when tuning the aforementioned subset. We then verify the most promising configurations on the original workload to measure their actual performance.

\begin{figure*}[t]
  \centering
  \hspace{1em}
  % \vspace{-2em}
  % \includegraphics[width=\textwidth]
  \includegraphics[scale=0.55]{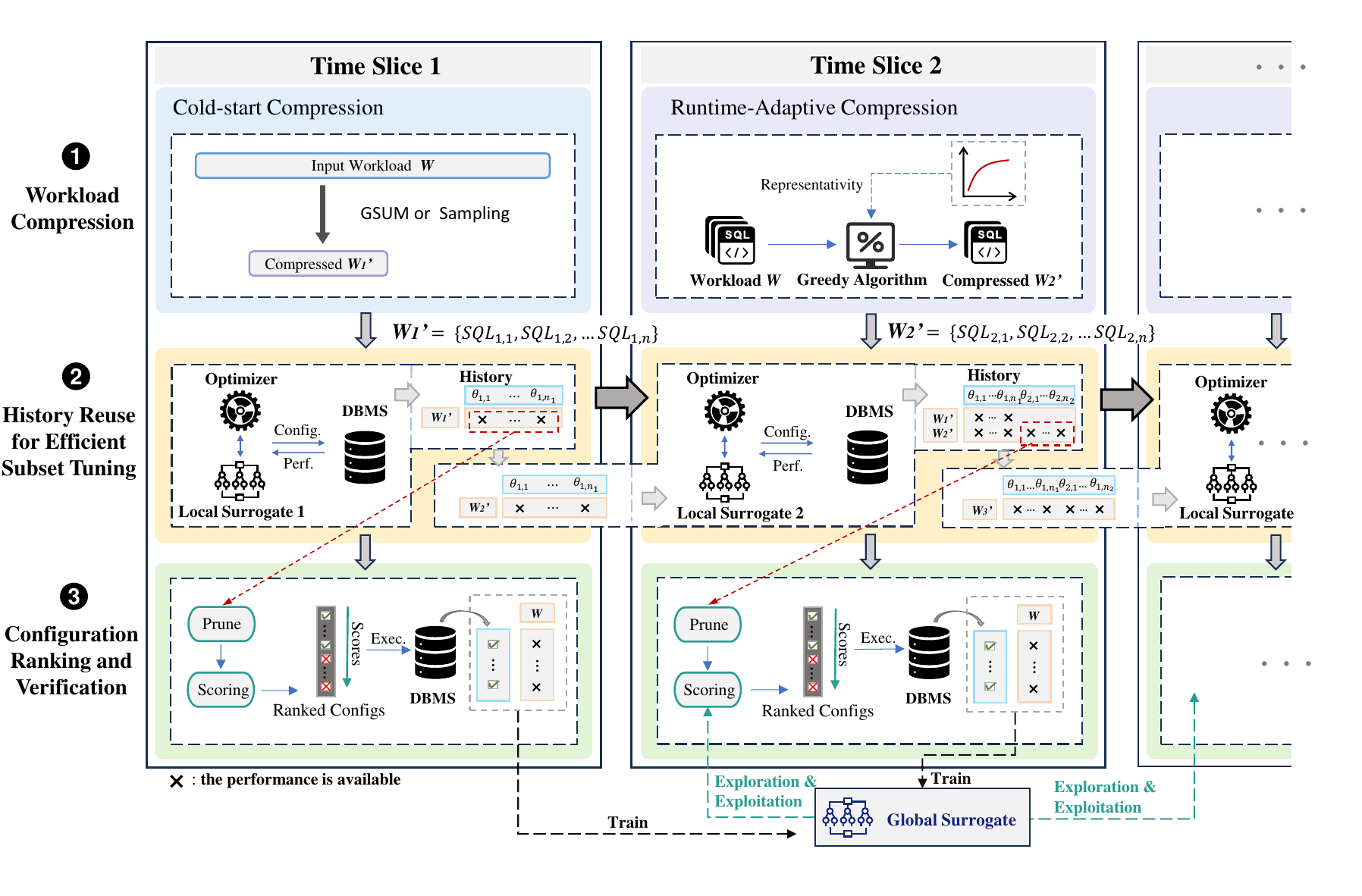}
  \vspace{-2em}
  \caption{\system \ Tuning Workflow}
  \label{fig: workflow}  
  \vspace{-1em}
\end{figure*}

\noindent \textit{\underline{Workflow.}} Figure \ref{fig: workflow} shows the tuning workflow. Instead of repeatedly replaying the entire workload or a fixed set of SQL queries, \system\ divides the tuning process into multiple time slices, each evaluating a small subset of queries. The tuning consists of a sequence of time slices, with each slice comprising three steps:\ding{182} \textbf{Workload Compression: }Given an input workload, \system\ uses a greedy algorithm driven by runtime statistics to compress the workload, aiming to maximize a custom \textit{representativity} metric (Section~\ref{sec: workload compression}). \textcolor{black}{Since there is no runtime profile at the beginning, \system \ uses existing methods like GSUM or random sampling to initialize the subset.} 
\ding{183} \textbf{Subset Tuning:} \system\ reuses its tuning history to initialize the local surrogate model for the current subset, thereby enabling efficient subset tuning that yields a series of configurations (Section \ref{sec: history-reuse}). \ding{184} \textbf{Configuration Verification:} \system \ uses heuristic rules and a hybrid scoring mechanism to identify the most promising configurations proposed in step \ding{183}, which it then evaluates on the entire workload to determine their actual performance (Section~\ref{sec: configuration}).

\section {Workload Compression}
\label{sec: workload compression}
In this section, we redefine the workload compression problem for knob tuning (Section \ref{sec: formulation}), introduce a runtime metric to measure the representativity of a SQL subset (Section \ref{sec: representative}), and optimizes this metric for runtime-adaptive workload compression (Section \ref{sec: runtime-adaptive}).

\subsection{Problem Formulation} \label{sec: formulation}

We first formulate the conventional workload compression problem and then redefine it within the context of knob tuning.

First, we formally define what is an original workload, a compressed workload and the corresponding compression ratio.

\begin{definition}[Original Workload]
\label{definition: original-workload}
\textit{Original Workload} is a multiset $\mathbf{W}=\{q_1, \dots, q_n\}$ consisting of $n$ SQL queries. Users' goal is to minimize the latency when executing this workload.
\end{definition}

\begin{definition}[Compressed Workload]
\label{definition: compressed-workle}
\textit{Compressed Workload} $\mathbf{W'}$ is a subset of $\mathbf{W}: \mathbf{W'}\subseteq \mathbf{W}$. Formally, $\mathbf{W'}=\{q_1, \dots, q_m\}$ where $q_i \in \mathbf{W}$ and $m\le n$.
\end{definition}

Next, since workload compression task should be constrained by a given budget $B$, we define a cost of each query as follows:

\begin{definition}[Query Cost]
\label{definition: tuning-lake} Each SQL query $q_i$ is associated with a non-negative cost $c(q_i)$, where $c(q_i)$ is a function that quantifies the cost a query introduces to the tuner $A$ in completing the tuning task. This cost could, for example, represent the number of indexable columns considered for index tuning, or the query execution time for knob tuning.
\end{definition}

\begin{definition}[Compression Ratio]
\label{definition: compression ratio}
\textit{Compression Ratio}, $\eta=1-\frac{c(W')}{c(W)}$, is the fraction of workload that has been pruned. 
\end{definition}

Let $C(\mathbf{W})$ be the execution time of workload $\mathbf{W}$ under the default configuration, and $C_{K(\textbf{W'}, A)}(\mathbf{W})$ be its execution time under the configuration $K(\mathbf{W'}, A)$ recommended by tuner $A$ for subset $\mathbf{W'}$.

The conventional workload compression problem is defined as follows: given a compression budget $B\geq 0$, construct a compressed workload $\mathbf{W'}\subseteq \mathbf{W}$ such that \cite{isum,2002,wred}:

\begin{itemize}
\item $\sum_{\mathbf{q}\in \mathbf{W'}}c(\mathbf{q})\le B$, i.e., the cost of the compressed workload is less than the budget.

\item $\mathbf{W'}=\mathop{\arg\max}\limits_{\mathbf{W'}\subseteq\mathbf{W}}C(\mathbf{W})-C_{K(\mathbf{W'},~A)}(\mathbf{W})$, i.e., the reduction in the execution time of $\mathbf{W}$ is maximized when using the configuration $K(\mathbf{W'}, A)$.
\end{itemize}

Existing methods on workload compression reduce tuning costs by identifying a subset of SQL queries to tune, which takes less time to execute and can serve as the representative of the original workload. {These approaches prioritize a reduction in tuning time (i.e., runtime efficiency) at the expense of the resulting configuration's performance} \cite{wred,isum,gsum,query2vec}.

\textcolor{black}{While our method produces a configuration that achieves better performance than tuning on the full workload given the same tuning time.} Given a time budget $t$, we \textit{redefine} the workload compression problem in the context of knob tuning as follows:

\qquad \qquad \textbf{maximize} \quad $C_{K(\mathbf{W}, ~A, ~t)}(\mathbf{W}) - C_{K(\mathbf{W'}, ~A, ~t)}(\mathbf{W})$
\\

\qquad \qquad \textbf{subject to} \quad $\sum_{\mathbf{q}\in \mathbf{W'}} c(\mathbf{q})\leq B,$\quad $\mathbf{W'}\subseteq \mathbf{W},$

\vspace{1em}

\noindent
where $K(\mathbf{W},A,t)$ is the configuration recommended by tuner $A$ for $\mathbf{W}$ within time budget $t$, and $c(\mathbf{q})$ is the execution time of query $\mathbf{q}$ under the default configuration. Following prior work \cite{2002,gsum,isum,wred}, workload compression must be highly efficient, avoiding expensive operations like query execution or computing complex statistics.

\textcolor{black}{There is an inherent trade-off between the quality of feedback and the associated evaluation cost in each tuning iteration. Traditional methods execute the entire workload to obtain feedback, producing high-quality results but incurring substantial overhead. In contrast, our method evaluates only a subset of the workload to reduce costs. Although this sacrifices some accuracy in each feedback iteration, it enables a greater number of tuning iterations within the same time constraints. As discussed in \textbf{M1} and \textbf{M2}, given the extremely large search space, the number of tuning iterations is insufficient to explore such a large space, and this is the main bottleneck of knob tuning. To address this challenge, we select a representative subset for tuning, trading off per-iteration feedback quality for an increased overall number of iterations, and finally achieving better performance than tuning the original workload under the same time budget. Moreover, we propose methods to mitigate the impact of reduced per-iteration quality as much as possible, which are discussed next.
}

Although $\mathbf{W'}$ allows faster convergence by evaluating more configurations due to reduced workload execution time, $K(\mathbf{W}, A, t)$ will eventually outperform $K(\mathbf{W'}, A, t)$ with a sufficiently large tuning time budget $t$. This happens because $\mathbf{W'}$ is just an approximation of $\mathbf{W}$, and the bias between them will eventually lead to the optimization stagnating in later stages (see Figure \ref{fig: show_random}).
To address this, instead of maintaining a fixed compression ratio $\eta$, we: (1) refine the SQL subset without changing $\eta$, (2) once the subset’s capacity is reached, decrease $\eta$ to include more queries. This strategy balances the efficiency of subset tuning with the thoroughness of full workload tuning as the process continues.

% \vspace{-1em}
\subsection{Representative Subset}
\label{sec: representative}

In this section, we introduce a \textit{representativity} metric to measure how closely a selected subset’s behavior aligns with the original workload in the context of knob tuning.

\noindent\textbf{What is a representative subset?} 
A representative subset is a small collection of queries whose performance accurately mirrors that of the full workload across different system settings. The goal is to preserve relative performance: if configuration A is faster than configuration B on the subset, it must also be faster on the full workload. This alignment is essential because system tuning relies on knowing whether one configuration is better than another, not on their absolute execution times. A truly representative subset ensures optimization decisions are based on this reliable ranking. This is why many statistical and other methods \cite{gsum,isum,wred,2002,2003} are inadequate—they fail to maintain this critical performance relationship across configurations.

\noindent \textbf{Representativity Metric Definition.} Before introducing representativity, we need to maintain a run history defined as follows.

\begin{definition}[Run History]
\label{definition: history}
\textit{Run history} $H$ is a two-dimensional table  recording each query’s execution time across all evaluated configurations. Specifically, $H[\textbf{q}, \theta]$ represents the execution time of $\textbf{q}$ under configuration $\theta$. 
\end{definition}

\example 
\normalfont Table \ref{tab:history} illustrates an example of run history. The execution time of the workload ${q_1, q_2, \dots, q_n}$ under configuration $\theta$, denoted by $H[{q_1, q_2, \dots, q_n}, \theta]$, is the sum of the individual query execution times: $H[{q_1}, \theta] + H[{q_2},\theta] + \dots + H[{q_n}, \theta]$.

\textcolor{black}{The run history is updated each time a query is executed.} Using this run history, we calculate \textit{representativity} based on concordant performance pairs \cite{opadviser,rover}. For two configurations, $\theta_1$ and $\theta_2$, and two workloads, $\textbf{W}$ and $\textbf{W'}$, a performance pair is \textit{concordant} if the ranking of $(H[\textbf{W}, \theta_1], H[\textbf{W}, \theta_2])$ matches that of $(H[\textbf{W'}, \theta_1], H[\textbf{W'}, \theta_2])$. Here, $H[\textbf{W}, \theta_1]$ denotes the execution time of workload $\textbf{W}$ under configuration $\theta_1$.

\begin{definition}[Representativity]
\label{definition: representativity}
\textit{Representativity} of a compressed workload $\mathbf{W'}$ to its original workload $\mathbf{W}$ can be computed as:
\end{definition}

\begin{equation}
\begin{aligned}
R(\mathbf{W'}, \mathbf{W}) = 
& \frac{2}{|H| \times (|H| - 1)} \sum_{j=1}^{|H|} \sum_{k=j+1}^{|H|} \bigg( \mathbf{1}(H[\theta_j, \mathbf{W}] \le H[\theta_k, \mathbf{W}]) 
\\
& \oplus \mathbf{1}(H[\theta_j, \mathbf{W'}] \le H[\theta_k, \mathbf{W'}]) \bigg).
\end{aligned}
\end{equation}
where $|H|$ is the number of configurations in $H$, and $\oplus$ is the exclusive-nor operator. \textcolor{black}{Essentially, \textit{representativity} is the ratio of concordant performance pairs between the two workload in the history $H$.}

\example
\normalfont Assume we have obtained the execution time of $\textbf{W}$ as (4, 5, 7) and $\textbf{W'}$ as (3, 2, 6) over three configurations $\theta_1$, $\theta_2$, and $\theta_3$, respectively. Then $R(\mathbf{W'}, \mathbf{W})$ is computed as follows:\\
\indent \textbf{1.} Pair the configurations in all possible combinations. We get $(\theta_1, \theta_2)$, $(\theta_1, \theta_3)$ and $(\theta_2, \theta_3)$.\\
\indent \textbf{2.} Judge the consistency of the performances of the two workloads on each configuration pair. We get $\mathbf{1}(4\le5) \odot \mathbf{1}(3\le2)=0, \mathbf{1}(4\le7)\odot \mathbf{1}(3\le6)=1$ and $ \mathbf{1}(5\le7)\odot \mathbf{1}(2\le6)=1$.\\
\indent \textbf{3.} Compute $R(\mathbf{W'}, \mathbf{W})=\frac{2\times (0+1+1)}{3\times2}=\frac{2}{3}$.

Representativity $R(\textbf{W'}, \textbf{W})$ ranges from $[0, 1]$. \textcolor{black}{A higher $R(\textbf{W'}, \textbf{W})$ indicates that $\textbf{W'}$ performs more similar to $\textbf{W}$ across different configurations, and thus $\textbf{W'}$ is more representative.} In practice, $R(\textbf{W'}, \textbf{W})$ typically falls within $(0.5, 1]$, since random performances yields $R(\textbf{W'}, \textbf{W})=0.5$. When $R(\textbf{W'}, \textbf{W})=1$, two workloads are equivalent for knob tuning.

\subsection{Runtime-Adaptive Compression}
\label{sec: runtime-adaptive}

\textcolor{black}{In this section, we demonstrate how to derive a representative subset of SQL queries from the evolving runtime profile by employing a greedy algorithm that optimizes the \textit{representativity} metric. Subsequently, we introduce the adaptive compression strategy.}

\noindent \textbf{Greedy Algorithm-based SQL Subset Selection.} We formalize the compression problem as follows:

\qquad \qquad \textbf{maximize} \quad $R(\mathbf{W'}, \mathbf{W})$
\\

\qquad \qquad \textbf{subject to} \quad $\sum_{\mathbf{q}\in \mathbf{W'}} c(\mathbf{q})\leq B,$\quad $\mathbf{W'}\subseteq \mathbf{W}$

% \vspace{1em}

\noindent Optimizing the set function in this formulation is NP-hard \cite{nphard}. However, we need to calculate an effective compressed workload with low overheads, otherwise we would lose the very purpose of workload compression in the first place \cite{2002}. Therefore, we adopt a greedy algorithm \cite{gsum,isum,fisher} that trades-off accuracy to optimize \textit{representativity} efficiently. And Feige \cite{feige} proved that, unless $P=NP$, no polynomial-time algorithm can achieve a better approximation ratio ($1 - 1/e$) than this greedy algorithm.

Instead of enumerating all possible query combinations and finding the one which maximizes \textit{representativity}, we loop over queries in $\textbf{W}$ time after time, and each time we add one query that maximizes the normalized marginal gain $\Delta(\mathbf{q}|\mathbf{W'}_{i-1})$ to the current compressed workload $\mathbf{W'}_i$. The marginal gain is defined as
\begin{equation}
\Delta(\mathbf{q}|\mathbf{W'})=\frac{R(\mathbf{W'}\cup\{\mathbf{q}\}, \textbf{W})-R(\mathbf{W'}, \textbf{W})}{c(\mathbf{q})}
\end{equation}

% $$\Delta(\mathbf{q}|\mathbf{W'})=\frac{R(\mathbf{W'}\cup\{\mathbf{q}\}, \textbf{W})-R(\mathbf{W'}, \textbf{W})}{c(\mathbf{q})}.$$
In other words, the algorithm greedily chooses the query with the best gain per unit of cost \cite{gsum}. 
% A $1-\frac{1}{e}$ approximation guarantee \cite{nphard} is achieved by it when optimizing objectives holding two attributes: (1) \textit{monotonicity} which means adding more samples cannot decrease the function value, and (2) \textit{submodularity} which means the marginal gain of adding a new element decreases as the set grows. It is apparent that \textit{representativity} holds these two attributes.

More importantly, adding a new query during knob tuning introduces execution overhead. We must execute this query under missing configurations to collect the cost data needed to bootstrap the surrogate models (Section \ref{sec: history-reuse}). Since this overhead varies significantly across queries, our compression process must account for the disparity. We quantify the additional overhead of a new query as follows:

\begin{definition}[Lacked History]
\label{definition: lacked-history}
Given a run history $H$ that records each query's performance across all proposed configurations (see Table \ref{tab:history}), $\#lacked\_history(\mathbf{q})$ is the number of configurations in $H$ for which performance on $\mathbf{q}$ is missing.
\end{definition}

We define the final marginal gain (before normalization) to simultaneously maximize representativity and minimize additional costs as follows:
\begin{equation}
\label{gain}
    \Delta(\mathbf{q}|\mathbf{W'})=\frac{R(\mathbf{W'}\cup\{\mathbf{q}\}, \textbf{W})-R(\mathbf{W'}, \textbf{W})}{c(\mathbf{q})}-\beta \times \#lacked\_history(\textbf{q})
\end{equation}

% $$\Delta(\mathbf{q}|\mathbf{W'})=\frac{R(\mathbf{W'}\cup\{\mathbf{q}\}, \textbf{W})-R(\mathbf{W'}, \textbf{W})}{c(\mathbf{q})}-\beta \times \#lacked\_history(\textbf{q}). $$

where $\beta$ serves as a hyperparameter that balances overhead and the marginal gain of adding a new query. As $\beta$ increases, queries with fewer missing performance records in existing configurations are more likely to be selected.

\noindent \textcolor{black}{\noindent \textbf{Adaptive Compression Strategy.} Workload compression occurs at the beginning of each time slice, leveraging an evolving runtime profile to continuously refine the subset. As more data becomes available, the subset becomes increasingly representative. Periodically updating the subset also prevents the optimization process from getting trapped in local optima, which can happen with a fixed subset. The compression ratio $\eta$ is dynamic and decreases to increase the subset size when optimization fails to find a better configuration within a time slice. This indicates that the subset may not be sufficiently representative, reaching its representativity limit. Although reducing $\eta$ increases overhead, it enhances subset representativity and enables more effective optimization.
}

% \vspace{-0.5em}
\section{History Reuse for Efficient Tuning}
\label{sec: history-reuse}
% Configuration Proposing
After workload compression, we get a newly selected SQL subset which is then frequently evaluated to guide the optimization. In this section, we first introduce the challenge when tuning different subsets in different time slices, then we discuss how we address this challenge to achieve efficient subset tuning.

\noindent \textbf{Challenge.}  Effective knob tuning depends on a well-trained surrogate model that accurately predicts a workload’s performance across various configurations. Existing methods maintain a single surrogate because they focus on a fixed workload \cite{gptuner,ea,llamatune}. In contrast, we tune different SQL subsets over time slices, necessitating a new surrogate for each subset since one surrogate cannot model multiple workloads. Bootstrapping a surrogate from scratch is costly, requiring numerous workload executions to gather training data. Although some transfer learning techniques enhance efficiency, they demand collecting thousands to tens of thousands of observations in advance \cite{restune,ottertune,qtune}, which is time-consuming and not universally applicable across different systems, hardware, and workloads. Moreover, transferred observations may not fit new subsets well, potentially misleading the optimization process.

\noindent \textbf{History Reuse for Surrogate Bootstrapping.} We leverage execution statistics for queries in the selected subset $\textbf{W'}$ from previous time slices, recorded in the tuning history $H$, to bootstrap the surrogate without expensive workload executions. \textcolor{black}{There are two scenarios: \textit{S1 (Complete History):} If every query in $\textbf{W'}$ has been executed for all configurations in $H$, we sum their execution times per configuration to determine the total execution time for $\textbf{W'}$ on those configurations. This data is then used to bootstrap the surrogate (Example 3). \textit{S2 (Incomplete History):} If some queries in $\textbf{W'}$ lack execution times for certain configurations, we execute these missing queries for those configurations (Example 4) and then aggregate as in S1. Although this incurs some costs, it is significantly cheaper than bootstrapping the surrogate from scratch. To account for these costs, we incorporate a penalty term in the marginal gain computation as discussed in Section \ref{sec: runtime-adaptive}.}

This method allows us to bootstrap the surrogate on the fly without costly initial workload executions and ensures the data accurately reflects the subset's performance. As optimization progresses, accumulating data enhances the surrogate's accuracy for subsequent time slices.

\noindent \textbf{Subset Tuning.}
After bootstrapping the \textit{surrogate}, we use the tuner to optimize the subset. In each iteration, a proposed configuration is evaluated on the subset to update its best performance, and the resulting performance metrics are sent to the tuner to guide subsequent optimizations.

\begin{table}[htbp]
  \centering
  \caption{A Toy Example of History}
  \vspace{-0.5em}
    \begin{tabular}{c|cccc}
    \toprule
    \multicolumn{1}{p{2.555em}|}{\textbf{ }} & $\theta_1$ & $\theta_2$ & $\theta_3$ \\
    \midrule
    \rowcolor[rgb]{ .851,  .851,  .851} $\mathbf{q}_1$  & $H[\mathbf{q}_1, \theta_1]$   & $H[\mathbf{q}_1, \theta_2]$   & $H[\mathbf{q}_1, \theta_3]$  \\
    $\mathbf{q}_2$  & $H[\mathbf{q}_2, \theta_1]$   &       & $H[\mathbf{q}_2, \theta_3]$  \\
    \rowcolor[rgb]{ .851,  .851,  .851} $\mathbf{q}_3$  & $H[\mathbf{q}_3, \theta_1]$   &       &     \\
    $\mathbf{q}_4$  & $H[\mathbf{q}_4, \theta_1]$   &   $H[\mathbf{q}_4, \theta_2]$    &  $H[\mathbf{q}_4, \theta_3]$   \\

    \bottomrule
    \end{tabular}%
  \label{tab:history}%
  \vspace{-0.7em}
\end{table}%

\example 
\normalfont Assume $\mathbf{W'}=\{\mathbf{q}_1, \mathbf{q}_4\}$ for the subsequent time slice and we have run history illustrated in Table \ref{tab:history}. The surrogate for the next time slice should be bootstrapped with the data:\\ $\{(\theta_1, H[\{\mathbf{q}_1, \mathbf{q}_4\}, \theta_1]), (\theta_2, H[\{\mathbf{q}_1, \mathbf{q}_4\}, \theta_2]), (\theta_3, H[\{\mathbf{q}_1, \mathbf{q}_4\}, \theta_3])\}$.

\example 
\normalfont Assume $\mathbf{W'}=\{\mathbf{q}_2, \mathbf{q}_3\}$ for the subsequent time slice and we have run history illustrated in Table \ref{tab:history} which lacks $H[\mathbf{q}_2, \theta_2]$, $H[\mathbf{q}_3, \theta_2]$, $H[\mathbf{q}_3, \theta_3]$. We need to deploy $\theta_2$ and run $\mathbf{q}_2, \mathbf{q}_3$, and  deploy $\theta_3$ and run $\mathbf{q}_3$.

\section{Configuration Pruning, Ranking, Verification}
\label{sec: configuration}

\textcolor{black}{After multiple subset tuning iterations, we identified and evaluated several configurations on the selected subset. However, our ultimate goal is to find configurations that perform well on the entire workload and report the real performance. To avoid exhaustive verification, we focus only on promising configurations by applying heuristic rules to eliminate unpromising ones, ranking the remaining options with a hybrid scoring mechanism, and selecting top configurations (e.g., 30\%) for verification on the entire workload.
}

\noindent\textcolor{black}{\textbf{Motivation.}
% In optimization problems with complex configuration spaces, 
}
\noindent\textcolor{black}{When selecting configurations for the entire workload, we face the exploration–exploitation dilemma \cite{dilemma}. \textit{Exploitation} chooses the best configuration based on current knowledge, including subset performance and model predictions. However, this can lead to suboptimal configurations, as a configuration that performs well on a subset may perform poorly on the full workload. Additionally, prediction models may be biased toward familiar configurations while underestimating unfamiliar ones. \textit{Exploration}, on the other hand, involves testing some unfamiliar configurations to discover unexpectedly high performers. Balancing exploration and exploitation is essential for achieving a global optimum. To address this, we propose a \textit{hybrid scoring mechanism} that effectively balances both strategies by scoring candidate configurations and selecting the top-ranked ones for further verification.
}

\noindent \textbf{Global Surrogate.} We maintain a global surrogate model, $\mathcal{RF}$, for scoring, trained on historical $(\theta, H[\mathbf{W}, \theta])$ pairs. It predicts performance and uncertainty estimates. We use a random forest regressor for its superior performance in knob tuning \cite{ea} and ability to quantify uncertainty \cite{smac}.

\begin{definition}[Uncertainty]
\label{definition: uncertainty}
\textit{Given an unlabeled configuration $\theta$ and a Random Forest $\mathcal{RF} = \{rf_1, \ldots, rf_n\}$ with $n$ estimators, the uncertainty $\Psi(\theta, \mathcal{RF})$ is the variance of their predictions for $\theta$.}
\end{definition}

\noindent \textbf{Exploitation.} \textcolor{black}{We approximate a candidate configuration $\theta$’s performance by combining its subset execution time $cost(\mathbf{W'})$ with the global surrogate $\mathcal{RF}$'s prediction.}
The \textit{predicted performance} (lower is better), $\hat{f}_\mathbf{W}(\theta)$, is formulated as follows:
% is composed of: (1) the scaled $\mathcal{RF}$' prediction and (2) its execution time $cost(\mathbf{W'})$ on the current SQL subset $\mathbf{W'}$ obtained during subset tuning:
\begin{equation}
    \hat{f}_\mathbf{W}(\theta) = -[(1 - \frac{|\mathbf{W'}|}{|\textbf{W}|})\mathcal{RF}(\theta) + cost(\mathbf{W'})].
\end{equation}

\noindent \textbf{Exploration.} Inspired by active learning \cite{ranked,hal,active}, we prioritize configurations that differ significantly from already labeled instances (i.e., $H[\mathbf{W}, \theta]$ is available) or those where the surrogate has low confidence in the prediction. We first introduce the definition of SetSimilarity and Uncertainty.

\begin{definition}[SetSimilarity]
\label{definition:similarity}
\textit{Given a labeled configuration set $\mathcal{D}$ and an unlabeled configuration $\theta$, SetSimilarity $\Phi(\theta, \mathcal{D})=\mathop{\max}\limits_{d \in \mathcal{D}}\phi(\theta, d)$, where $\phi$ is the similarity function.}
\end{definition}

We use the Gower distance $D(x,y)$, which measures the distance between two data points with mixed types of variables (numerical and categorical) \cite{gower}, to define the similarity function $\phi$:

\vspace{-0.5em}
$$
\phi(x,y)=\frac{1}{1+D(x,y)},
$$
where 
\vspace{-0.5em}
$$
D(x,y)=\frac{1}{n}\sum_{i=1}^n d_i(x,y),
$$
and for numerical variables:
\vspace{-0.5em}
$$
d_i(x,y)=\frac{|x_i-y_i|}{\max(x_i)-\min(x_i)},
$$
and for categorical variables:
\vspace{-0.5em}
$$
d_i(x,y)=
\begin{cases}
0, & \text{if } x_i=y_i \\
1, & \text{if } x_i\neq y_i
\end{cases}.
$$

\textcolor{black}{We use $1-\Phi(\theta, \mathcal{D})$ to give high scores to instances that do not share much similarity with already labeled documents (diversity).}

We use the verification ratio $\alpha$, the proportion of labeled to proposed configurations, to balance diversity and uncertainty prioritization. The \textit{exploration potential} of $\theta$ is defined as:

\begin{equation}
    g(\theta)=\alpha (1-\Phi(\theta, \mathcal{D}))+(1-\alpha)\Psi(\theta, \mathcal{RF})
\end{equation}

\noindent\textcolor{black}{\textbf{Hybrid Scoring Mechanism.} In each time slice, we randomly choose to either \textit{exploit} ($\hat{f}_\mathbf{W}(\theta)$) or \textit{explore} ($g(\theta)$) a configuration $\theta$. We select \textit{exploitation} with probability $1-\eta$ (subset volume), reflecting our current knowledge of $\theta$. The more we know about a configuration, the more likely we are to exploit it:
}

\begin{equation}
    S(\theta) = \begin{cases}
 \hat{f}_\mathbf{W}(\theta), & \text{with probability } 1-\eta\\
g(\theta), & \text{with probability } \eta
 \end{cases}
\end{equation}

\noindent \textcolor{black}{
\textbf{Configuration Pruning and Selection.} In each time slice, when $\hat{f}_\mathbf{W}$ is selected, configurations worse than the default are pruned. When $g$ is selected, configurations performing 1.2 times worse than the default are discarded. The remaining configurations are scored by the corresponding function, and the top-scoring ones are selected for verification. In the first time slice, due to the lack of labeled data, we simply discard configurations worse than the default and rank the remaining ones based on their performance on the compressed workload.}

\noindent \textbf{Verification.} 
To verify the selected configurations on the original workload $\textbf{W}$, we deploy them to the database and execute only the remaining subset $\mathbf{W}-\mathbf{W'}$, since $\mathbf{W'}$ was already evaluated during tuning. The execution results update the global surrogate, and \system \ outputs the best-performing configuration.

\section{Experiments}

\subsection{Experimental Setup}
\label{sec: setup}
\noindent \textbf{Workloads.}  
% In this work, we only focus on OLAP workloads which require variable-length workload execution periods to execute complex queries, whereas OLTP workloads are excluded because they are often evaluated with fixed workload execution periods.
\textcolor{black}{We focus exclusively on OLAP workloads, as OLTP workloads are typically evaluated over fixed intervals, making workload compression inapplicable.} Our experiments utilize three well-known database benchmarks: TPC-DS, JOB \cite{job}, and TPC-H. \textcolor{black}{Since TPC-DS is unsuitable for knob tuning \cite{ea}, we exclude templates with execution times significantly longer than others, following \cite{ds-1, ds-2, ds-3}.} For TPC-H, we use two variants: TPC-H and TPC-H×10, the latter of which includes 10 instances generated with different random seeds per template. Table \ref{tab:workload} summarizes the used workloads.

% \vspace{-0.5em}
\setlength{\tabcolsep}{3.6pt}
\begin{table}[htbp]
  \centering
  \caption{Summary of Workloads}
  \vspace{-0.5em}
    \begin{tabular}{ccccc}
    \toprule
    \textbf{Workload} & \textbf{Queries} & \textbf{Templates} & \textbf{Tables} & \textbf{Columns} \\
    \midrule
    \rowcolor[rgb]{ .85,  .85,  .85} TPC-DS* (sf=1) & 88    & 88    & 24    & 237 \\
    JOB (5.2GB) & 113   & 113   & 21    & 38 \\
    \rowcolor[rgb]{ .85,  .85,  .85} TPC-H (sf=10) & 22    & 22    & 8     & 55 \\
    TPC-H×10 (sf=10) & 220   & 22    & 8     & 55 \\
    \bottomrule
    \end{tabular}%
    
    \footnotesize
    \raggedright{\quad\textcolor{black}{*: template 1, 4, 6, 11, 14, 23, 24, 39, 74, 81, and 95 are removed.}}
  \label{tab:workload}%
\end{table}%

% \vspace{-1em}

\noindent \textbf{Hardware.} All experiments are conducted on (C1) a virtual machine with 32 vCPU and 60GB of RAM on a private server with an AMD EPYC 9654 96-Core Processor, or (C2) Alibaba Cloud Platform with an ecs.e-c1m4.xlarge instance with 4 vCPU and 16 GB of RAM.

\begin{figure*}[t]
    \centering
    \includegraphics[scale=0.43]{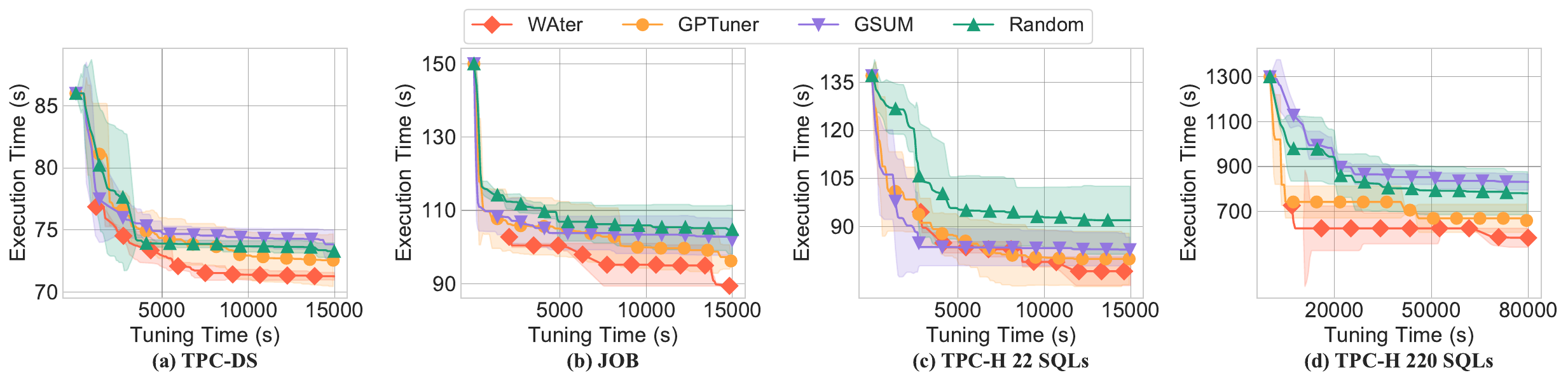}
    \vspace{-2em}
    \caption{Performance on different benchmarks (bottom-left is better)}
    \label{fig:main_experiment}
\end{figure*}

\begin{figure*}[t]
    \centering
    \includegraphics[scale=0.43]{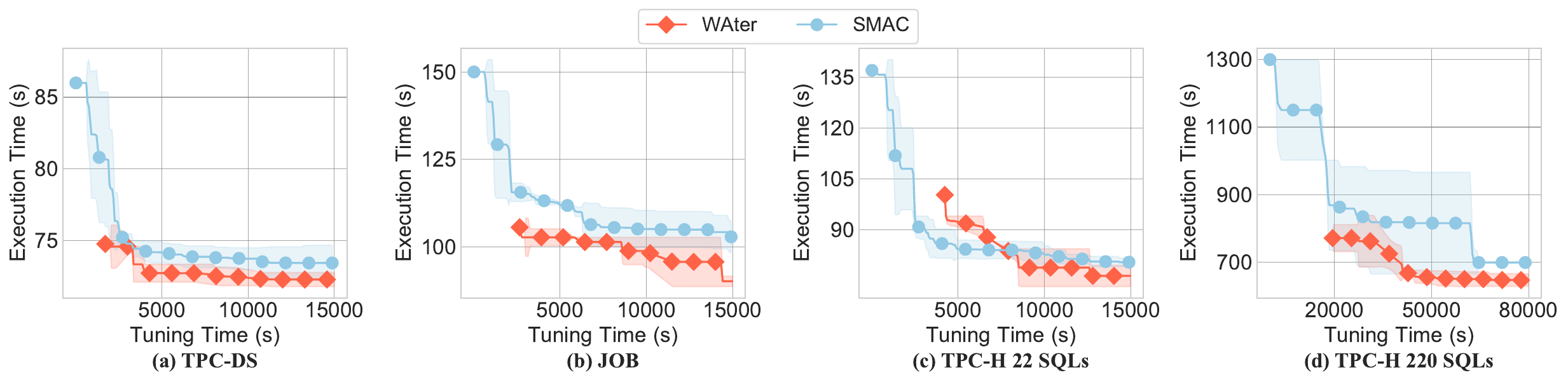}
    \vspace{-2em}
    \caption{Performance on different benchmarks (SMAC-based) (bottom-left is better)}
    \label{fig:smac}
\end{figure*}

\begin{figure}
    \centering
    \includegraphics[scale=0.39]{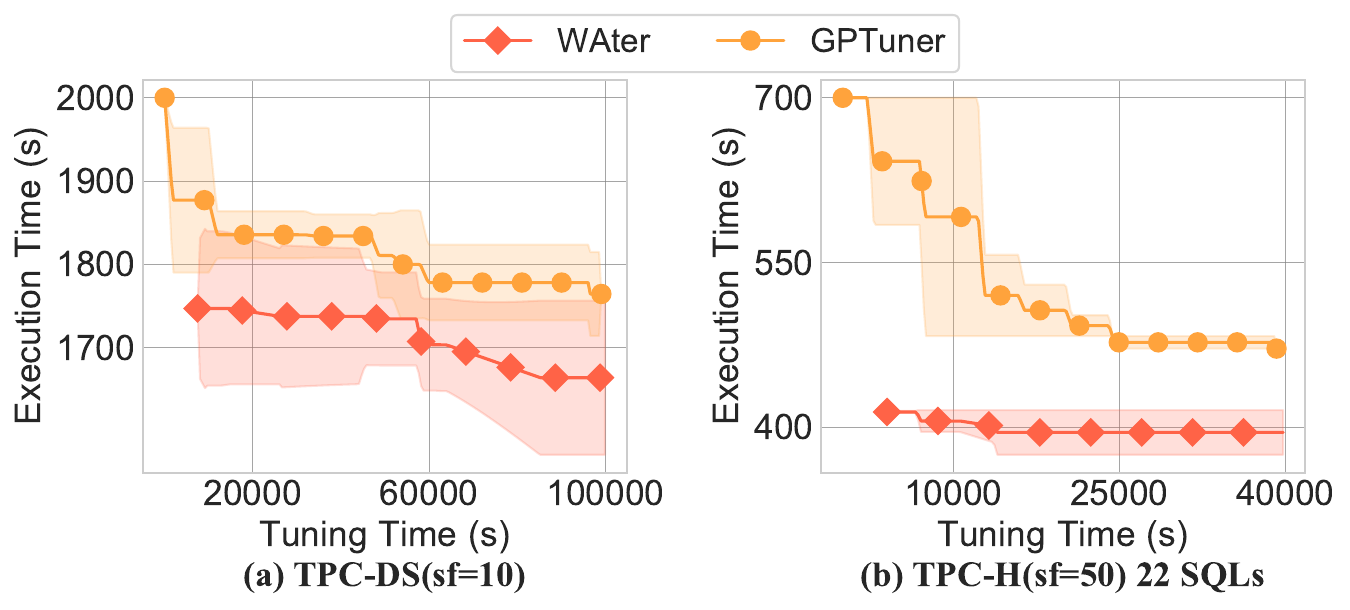}
    \vspace{-2em}
    \caption{Performance under different scale factors}
    \vspace{-1em}
    \label{fig:different_scale_factor}
\end{figure}

\begin{figure}
    \centering
    \includegraphics[scale=0.39]{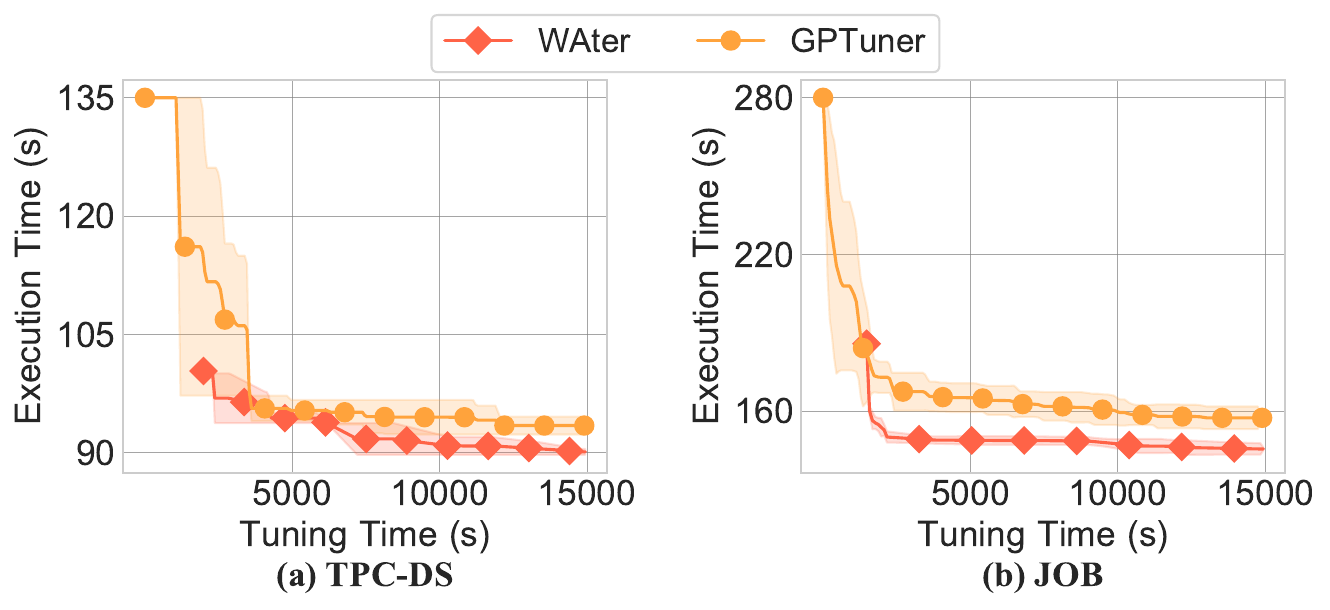}
    \vspace{-2em}
    \caption{Performance on different machine}
    \vspace{-2em}
    \label{fig:different_machine}
\end{figure}

\begin{figure*}[!thb]
    \begin{minipage}{0.23\textwidth}
        \begin{figure}[H]
            \centering
            \includegraphics[scale=0.38]{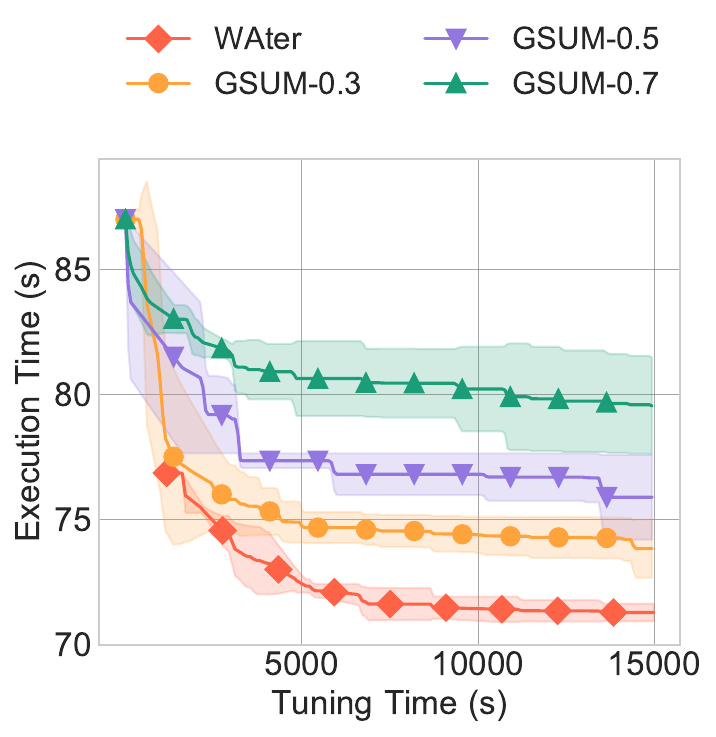}
            \vspace{-1.8em}
            \caption{GSUM with Different Compression Ratio}
            \label{fig: different_init_ratio}
        \end{figure}
    \end{minipage}
    \hfill
    \begin{minipage}{0.76\textwidth}
        \begin{figure}[H]
            \centering
            \includegraphics[scale=0.4]{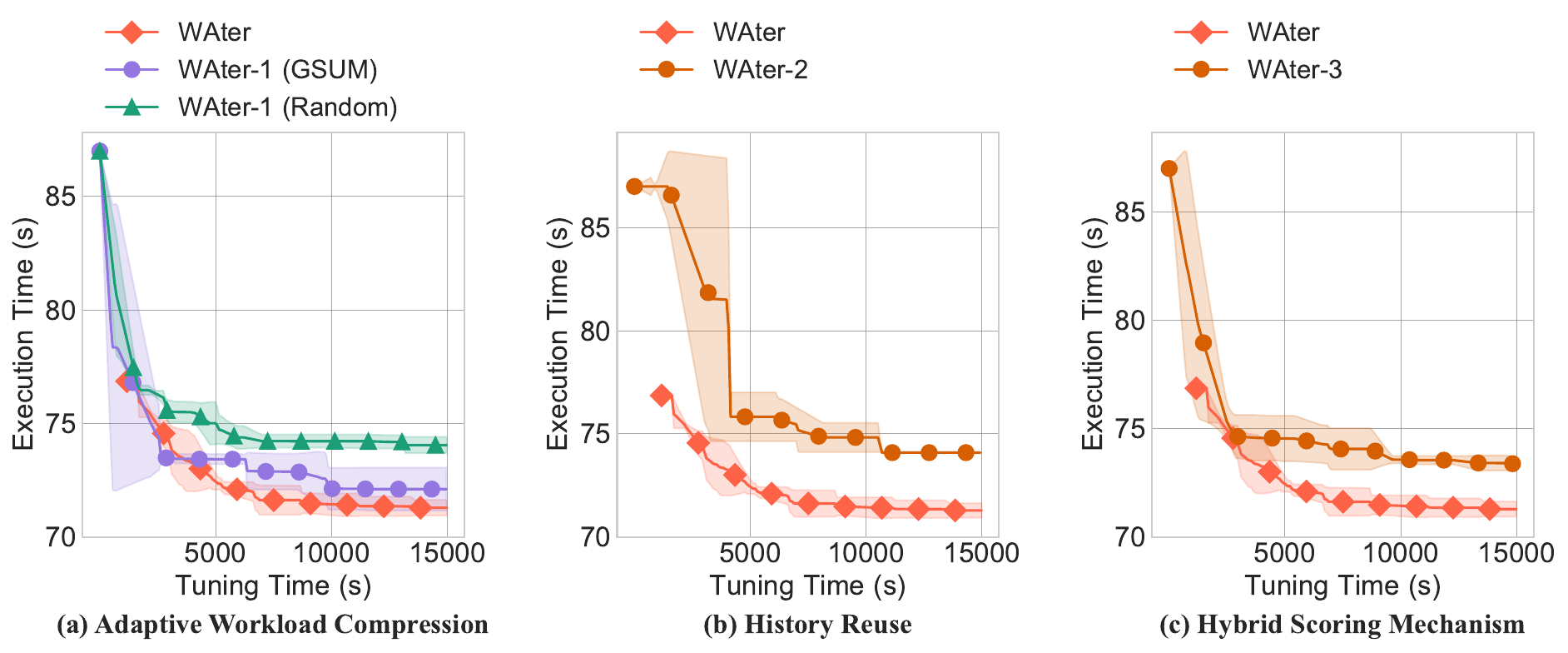}
            \vspace{-1em}
            \caption{Ablation study of \system \ on TPC-DS (bottom-left is better)}
            \label{fig: abalation}
        \end{figure}
    \end{minipage}%
\end{figure*}

\noindent \textbf{Adopted Tuners.} \system~is a generic optimization framework that enhances the tuning efficiency of existing tuners. We integrate it with \textbf{SMAC} \cite{smac3}, which recent evaluations \cite{ea} show outperforms eight state-of-the-art DBMS tuners, and with \textbf{\textsc{GPTuner}} \cite{gptuner}, which leverages domain knowledge for knob tuning. We utilize the open-source GPTuner code, updating its knowledge based on hardware, and implement SMAC using the SMAC3 \cite{smac3} library.

\noindent \textcolor{black}{\textbf{Baselines.} We compare \system \ with the following baselines:} \textbf{1. Original.} Utilizes the vanilla tuner (SMAC or \textsc{GPTuner}) to optimize the entire workload, highlighting \system's advantages. \textbf{2. GSUM \cite{gsum}.} A state-of-the-art workload compression method that maximizes both \textit{coverage} and \textit{representativity} as described in Section \ref{sec: background workload}. \textbf{3. Random.} Selects SQLs uniformly at random. Both \textbf{GSUM} and \textbf{Random} are static pre-processing techniques applied initially to obtain a subset for tuning. If a configuration outperforms the default on this subset, it is immediately evaluated on the entire workload. The compression ratios for \textbf{GSUM} and \textbf{Random} are set to be the same as \system's initial compression ratio by default.

\noindent \textbf{\system \ Implementation.} We implement \system \ in Python3 on top of the two tuners. The global surrogate uses scikit-learn’s RandomForestRegressor \cite{sklearn} with default parameters. The compression ratio $\eta$ starts at 0.75 and decreases by 0.1 if no better configuration is found within a time slice. In Equation \ref{gain}, $\beta$ is set to 0.1. In each time slice, 20 valid configurations are proposed during subset tuning, with 25\% (verification ratio $\alpha$) evaluated on the entire workload. For the initial time slice without runtime history, \textcolor{black}{we employ Latin Hypercube Sampling (LHS) \cite{lhs}, a space-filling sampling strategy, to generate ten samples for surrogate initialization, following previous works \cite{ea,ituned,llamatune,gptuner-award}.} \textcolor{black}{We use GSUM to select the initial subset.}

\noindent \textbf{Tuning Settings.} We conduct experiments with PostgreSQL v14.9, tuning 57 knobs from \textsc{GPTuner}'s open-source repository \cite{gptuner-code}. For each method, we perform three tuning sessions and report the average best performance (over the entire workload) with a solid line and $[5\%, 95\%]$ confidence interval shaded in the same color \cite{llamatune}. \textcolor{black}{Following \cite{dbbert,udo,unif}}, we use total workload execution time as the performance metric. Each method undergoes at least 100 tuning iterations, with the first 10 generated randomly using LHS \cite{lhs}, following previous works \cite{ea,ituned,ottertune,gptuner}. For failed or long-running configurations (those causing DBMS crashes or taking more than twice the execution time of the default), we assigned twice the default performance to prevent scaling issues \cite{inquiry}.

\noindent \textbf{Evaluation Metrics.} Following LlamaTune \cite{llamatune}, we use two metrics to evaluate \system: \textcolor{black}{\textbf{\textit{final performance improvement}}} (i.e., execution time reduction) and relative \textcolor{black}{\textbf{\textit{time-to-optimal speedup}}}, which reports the earliest time at which \system \ has found a better-performing configuration
compared to the baseline optimal, as well as the relative speedup.

% \vspace{-1em}
\subsection{Performance Comparison}\label{subsec:perf-comparison}
\noindent \textbf{End-to-end Comparison.} Figure \ref{fig:main_experiment} compares \system\ (integrated with \textsc{GPTuner}) against baselines. The initial gap in the red line reflects cold-start compression and subset tuning times during the first time slice.
Compared to \textbf{Original}, \system \ achieves the best performance identified by \textsc{GPTuner} $\mathbf{4.2\times}$ faster across all four workloads on average. Specifically, \system \ delivers time-to-optimal speedups of $\mathbf{2.5}\times$ for TPC-DS, $\mathbf{2.1}\times$ for JOB, and $\mathbf{11.0}\times$ for TPC-H$\times 10$, thanks to improved runtime efficiency. In terms of final results, \system \ reduces execution time by an average of $\mathbf{39.1\%}$ compared to the default and is $\mathbf{6.4\%}$ faster than \textsc{GPTuner}'s best. \textcolor{black}{\system's advantage over GPTuner on the TPC-H benchmark is initially modest, a result of two key factors. First, with only 22 queries, any small subset struggles to capture the diverse performance characteristics of the full workload. Second, the workload's shorter execution time lessens the overall impact of runtime efficiency gains. However, \system's dynamic strategy proves crucial in the long run. By continuously refining its query subset, it avoids the performance plateaus that cause static methods like GSUM and Random to stagnate, ultimately allowing it to find superior configurations in later stages.}

% shows only minor advantages over \textsc{GPTuner} on TPC-H, likely due to two factors: (1) With only 22 queries, any small subset struggles to capture the diverse performance characteristics of the full workload, and (2) the relatively short execution time of TPC-H reduces the impact of runtime efficiency. However, \system \ outperforms \textsc{GPTuner} in later stages, likely due to continuous subset refinement, preventing optimization stagnation as seen with \textbf{GSUM} and \textbf{Random}.

While \textbf{GSUM} and \textbf{Random} find interesting configurations early on, their optimization stagnates, ultimately failing to outperform \textsc{GPTuner} on average. Despite being a generic compression framework, \textbf{GSUM} does not always surpass random sampling in knob tuning, as its features are not specifically designed for this task and may lead to suboptimal compression. 
\textbf{Random} typically produces wider shadows in the figure, particularly for workloads with fewer SQLs (e.g., TPC-H), indicating greater instability. \textcolor{black}{To ensure fairness, we also compare \textbf{GSUM} under modified compression ratios with \system on TPC-DS. As shown in Figure \ref{fig: different_init_ratio}, \textbf{GSUM} underperforms \system~ on all compression ratios of 0.3, 0.5, and 0.7.}

% \vspace{-1em}
\subsection{Robustness Study}
\noindent \textbf{Different Optimizer.} To demonstrate \system's versatility with different optimizers, we replace the optimizer with SMAC. As shown in Figure \ref{fig:smac}, \system \ outperforms vanilla SMAC across all four workloads, achieving a $\mathbf{37.5\%}$ mean reduction in execution time compared to the default and $\mathbf{6.6\%}$ less time than SMAC’s best configuration. Additionally, \system \ provides a $\mathbf{3.1}\times$ time-to-optimal speedup on average. We achieve speedups of $\mathbf{3.8}\times$ and $\mathbf{5.3}\times$ on TPC-DS and JOB, reducing execution time by $\mathbf{15.9\%}$ and $\mathbf{40.0\%}$, respectively. While TPC-H remains a challenge for \system, it initially lags but ultimately outperforms the vanilla optimizer as the subset evolves.

\noindent \textbf{Different Data Size.} We study \system's scalability across different database sizes by varying the scale factor of TPC-H from 10 to 50 and TPC-DS from 1 to 10. As shown in Figure \ref{fig:different_scale_factor}, compared to \textsc{GPTuner}, \system \ finds better configurations in much less time. For both of the workloads, \system \ finds better configurations than the optima of \textsc{GPTuner} at the very beginning, achieving time-to-optimal speedups of $\mathbf{12.9}\times$ and $\mathbf{9.8}\times$ for TPC-DS and TPC-H. In the end, \system \ achieves execution times which are $\mathbf{16.8\%}$ and $\mathbf{43.5\%}$ less than default and $\mathbf{5.7\%}$ and $\mathbf{16.2\%}$ less than \textsc{GPTuner} on TPC-DS and TPC-H respectively. \textcolor{black}{This demonstrates that as the cost of a single evaluation increases, the benefit of \system's runtime efficiency becomes overwhelmingly significant, allowing it to explore many more configurations in the same time budget.}

\noindent \textbf{Different Hardware.} We switch from hardware C1 to C2, which has significantly fewer CPU cores and less RAM. This change makes optimization more challenging because reduced resources increase the complexity of modeling the relationship between configurations and DBMS performance, revealing more bottlenecks and shrinking the feasible region \cite{gptuner}. We exclude TPC-H experiments as they frequently cause system crashes on C2. Figure \ref{fig:different_machine} shows that compared to \textsc{GPTuner}, \system\ finds better configurations in fewer iterations for both workloads. For JOB, \system\ identifies a configuration superior to \textsc{GPTuner}'s best on the first attempt ($\mathbf{7.9}\times$ speedup) and ultimately achieves a workload execution time $\mathbf{7.6\%}$ less than \textsc{GPTuner}'s best. For TPC-DS, \system\ achieves a $\mathbf{1.9}\times$ time-to-optimal speedup and reduces execution time by $\mathbf{3.6\%}$.

\begin{figure}
    \centering
    \includegraphics[scale=0.37]{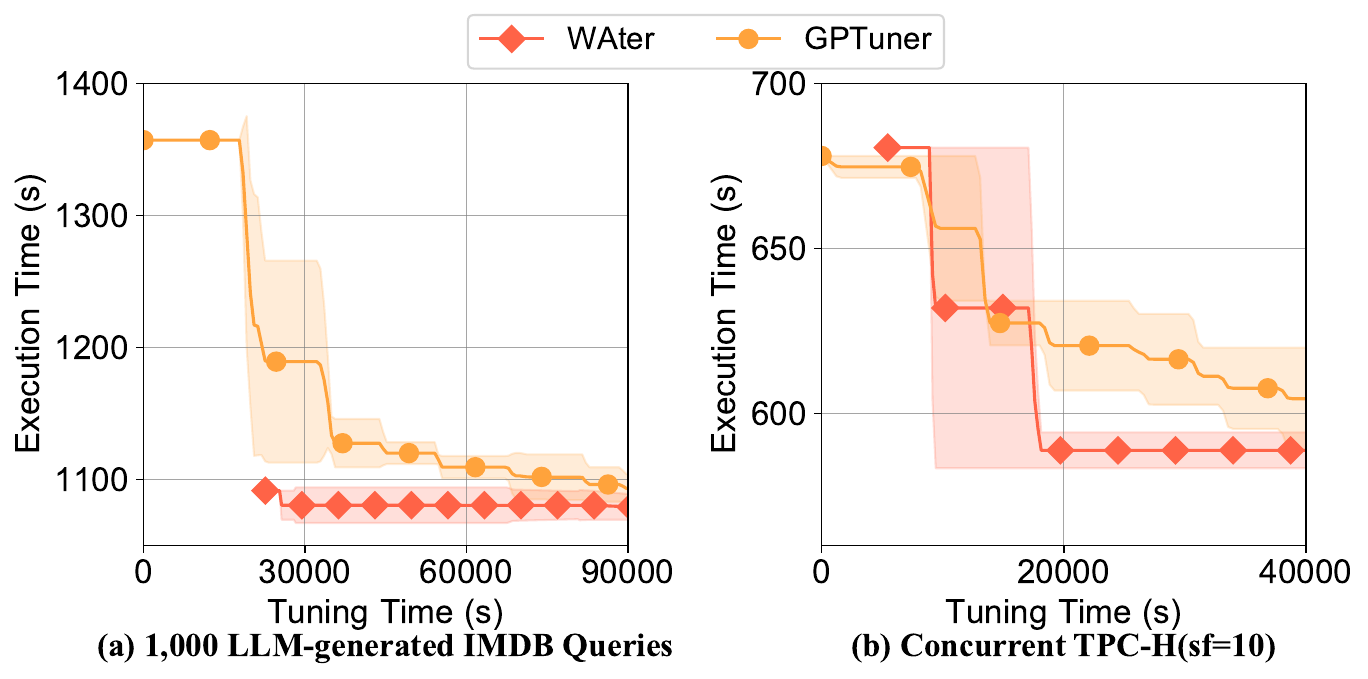}
    \vspace{-2em}
    \caption{Robustness under under concurrent and LLM-generated workloads}
    \vspace{-2em}
    \label{fig:new_workloads}
\end{figure}

\noindent \textbf{Expanded LLM-generated Workload.} Aligning with the recent trend of applying LLMs to SQL processing tasks \cite{sqlstorm,quite,dialect,sqlbarber, sqlbarber-demo, resq}, we evaluate \system\ on a workload of 1,000 IMDB-based queries generated by SQLStorm \cite{sqlstorm}, an LLM-driven tool for synthesizing SQL workloads. Figure \ref{fig:new_workloads} (a) demonstrates that \system\ achieves a $\mathbf{3.95}\times$ speedup in time-to-optimal convergence over \textsc{GPTuner}.

\noindent \textbf{Concurrent Workload.} To evaluate concurrent execution, we generate 50 groups of randomly sampled TPC-H queries. These groups range in size from two to six queries, with exactly 10 groups created for each size. Queries within the same group are treated as a single, indivisible unit and executed simultaneously. As shown in Figure \ref{fig:new_workloads}(b), \system\ achieves a $\mathbf{2.21}\times$ time-to-optimal speedup and reduces execution time by $\mathbf{2.59\%}$.

% As shown in Figure \ref{fig:new_workloads} (a), \system\ achieves a $\mathbf{3.95}\times$ time-to-optimal speedup compared to \textsc{GPTuner}.

% To evaluate concurrent execution, we generate 50 groups of randomly sampled TPC-H queries. These groups range in size from two to six queries, with exactly 10 groups created for each size. Queries in the same group are considered as minimal unit and executed concurrently.

% We construct a concurrent TPC-H workload by organizing queries into groups for simultaneous execution. We randomly sample queries to form group sizes ranging from two to six. By generating 10 distinct groups for each size, we create a total of 50 concurrent groups.

% \vspace{-0.5em}
\subsection{Ablation Study}
\noindent \textbf{Effect of Adaptive Workload Compression.}
We evaluate our adaptive workload compression framework and the algorithm from Section \ref{sec: workload compression} by keeping the subset fixed across all time slices. Using the same subsets as \textbf{GSUM} and \textbf{Random}, denoted ``\system -1 (GSUM)'' and ``\system -1 (Random)'', Figure \ref{fig: abalation}(a) shows that \system\ outperforms both, achieving speedups of $\mathbf{2.0\times}$ and $\mathbf{4.3\times}$, and reducing execution time by $\mathbf{1.1\%}$ and $\mathbf{3.6\%}$, respectively.

\noindent \textbf{Effect of History Reuse.}
To assess \textit{History Reuse for Efficient Subset Tuning} (Section \ref{sec: history-reuse}), we use LHS \cite{lhs} to randomly sample and evaluate configurations to bootstrap the surrogate in each time slice, which is referred to as ``\system -2''. As shown in Figure \ref{fig: abalation}(b), \system -2 stagnates early in optimization, while \system\ achieves an additional $\mathbf{3.8\%}$ reduction in execution time and a $\mathbf{3.5\times}$ speedup. The result is due to \system -2’s initialization overhead and undertrained surrogates from limited observations.

\noindent \textbf{Effect of Hybrid Scoring Mechanism.}
To demonstrate the hybrid scoring mechanism’s effectiveness (Section \ref{sec: configuration}), we replace it with a scoring method based solely on subset performance, denoted ``\system -3''. Figure \ref{fig: abalation}(c) shows that \system\ achieves a $\mathbf{4.1\times}$ speedup and reduces execution time by $\mathbf{3.1\%}$ compared to \system -3. This is because \system -3\ cannot reliably identify configurations that perform well across the entire workload, since configurations that perform well on the subset do not necessarily also perform well across the entire workload.

\subsection{Cost Analysis}
\label{sec: cost analysis}
\begin{figure}[h]
    \centering
    \subfloat[TPC-H (sf=10)]{
        \includegraphics[scale=0.34]{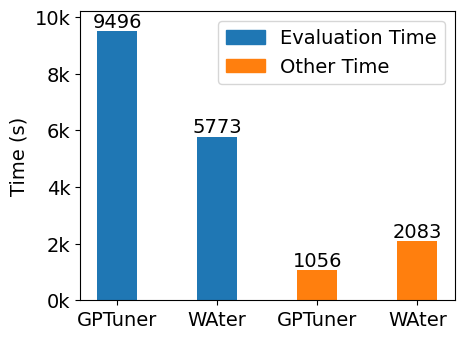}
        \label{fig: intro-compare-tpch}
    }
    \subfloat[TPC-H (sf=50)]{
        \includegraphics[scale=0.34]{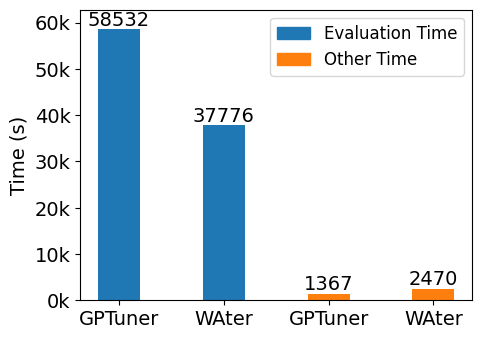}
        \label{fig: intro-compare-tpcc}
    }
    \vspace{-1em}
    \caption{Cost Analysis}
    \label{fig: cost analysis}
    \vspace{-1em}
\end{figure}

We divide the tuning time into two parts: (1) Evaluation Time, the duration spent executing queries, and (2) Other Time, covering tuner's overhead, algorithmic overhead and so on. Figure \ref{fig: cost analysis} shows the time spent in both categories during 100 tuning iterations for TPC-H (sf=10) and TPC-H (sf=50) using \system\ and \textsc{GPTuner}. \system \ reduces the overall tuning time by $\mathbf{25.5\%}$ and $\mathbf{32.8\%}$ for TPC-H (sf=10) and TPC-H (sf=50), respectively, compared to \textsc{GPTuner}, primarily due to a reduction in ``Evaluation Time''. Although \system\ incurs more ``Other Time'' due to additional overhead (e.g., model training, more configuration deployments), the large decrease in ``Evaluation Time'' more than offsets this. \system's advantage is particularly significant for workloads with a large "Evaluation Time," as it can substantially reduce this component. In contrast, "Other Time" remains unaffected by workload size and stays constant. \textcolor{black}{This explains why \system\ performs better on workloads with larger scale factors (Figure \ref{fig:different_scale_factor}). In real-world production environments, evaluation times of OLAP workloads are typically much longer than those presented in our experiments \cite{why-tpc,analytics_benchmark,snow_flake}, where \system\ demonstrates even greater potential.}

\section{Conclusion}
In this paper, we introduced \system, a workload-adaptive tuning system that shifts the focus of automated database knob tuning from sample efficiency to runtime efficiency. Instead of evaluating configurations against a full, resource-intensive workload, \system\ dramatically reduces tuning costs by evaluating small, dynamically refined query subsets within discrete time slices. Extensive evaluations across multiple OLAP benchmarks demonstrate that \system\ identifies near-optimal configurations an average of \textbf{4.2x} faster than state-of-the-art baselines, achieving up to \textbf{16.2\%} better final performance. Furthermore, \system\ maintains strong robustness across diverse hardware environments, optimizers, database scales, workload execution modes, and workload types.

\bibliographystyle{ACM-Reference-Format}
\bibliography{sample}

\end{document}